\documentclass[preprint,superscriptaddress,showpacs,aps,pre,floatfix,longbibliography]{revtex4-1}
\usepackage[utf8]{inputenc} 
\usepackage[T1]{fontenc}    
\usepackage{booktabs}       
\usepackage{amsfonts}       
\usepackage{nicefrac}       
\usepackage{microtype}      
\usepackage{amssymb,amsthm}
\usepackage{amsmath,bm}
\usepackage{graphicx,epstopdf}
\usepackage{multirow}
\usepackage{subfigure}
\usepackage{setspace}
\usepackage{wrapfig}
\usepackage{comment}
\usepackage{listings}
\usepackage{amssymb,dsfont}
\usepackage{leftidx}
\usepackage{bigdelim}
\usepackage{mathrsfs}
\usepackage{blkarray,adjustbox}
\usepackage{lineno}
\usepackage{enumitem}
\usepackage{mathpazo}
\usepackage{rotating}
\usepackage{soul}
\usepackage[dvipsnames]{xcolor}
\usepackage{xspace}
\usepackage{float}
\usepackage{titlesec} 
\usepackage[pdftex, pdftitle={Article},colorlinks, pdfauthor={Author}]{hyperref}
\usepackage[nameinlink,capitalise]{cleveref}
\usepackage{array, makecell, cellspace}
\hypersetup{colorlinks={true},linkcolor={darkgray},citecolor=ForestGreen}

\definecolor{reddish}{HTML}{FBB4AE}
\definecolor{blueish}{HTML}{B3CDE3}
\definecolor{magentish}{HTML}{FF00AA}
\definecolor{greenish}{HTML}{a1d99b}

\begin{document}

\title{The hierarchical organization of urban social overlap: How shared social spaces diverge from physical mobility}

\author{Bibandhan Poudyal}
\affiliation{Department of Physics \& Astronomy, University of Rochester, Rochester, NY, USA}
\author{Mariana Macedo}
\affiliation{Department of Computer Science, Constructor University, Bremen, Germany}
\author{Alexandre G. Evsukoff}
\affiliation{COPPE, Federal University of Rio de Janeiro, Rio de Janeiro, RJ, Brazil}
\email{alexandre.evsukoff@coc.ufrj.br}
\author{Ronaldo Menezes}
\email[Correspondence email address: ]{r.menezes@exeter.ac.uk}
\affiliation{BioComplex Laboratory, Department of Computer Science, University of Exeter, UK}
\author{Gourab Ghoshal}
\email[Correspondence email address: ]{gghoshal@pas.rochester.edu}
\affiliation{Department of Physics \& Astronomy, University of Rochester, Rochester, NY, USA}

\begin{abstract}
\begin{singlespace}
Cities are commonly characterized through population density, infrastructure, and mobility flows. These quantities describe where people are and how they move, but movement does not uniquely determine where populations repeatedly encounter one another or whether different locations participate in the same social environments. Here, we construct co-presence overlap networks across 13 Brazilian cities to distinguish these dimensions of urban organization. We introduce two complementary measures: colocation, which quantifies recurrent co-presence activity within individual locations, and co-connectedness, which measures the extent to which locations share the same co-present pairs. We find that the spatial concentration of colocation differs substantially from that of mobility, producing social centers that can differ in both number and location from mobility centers. At the network level, co-connectedness exhibits a non-random hierarchical organization in which locations preferentially share recurrent pairs with locations of similar activity. This hierarchy is systematically weaker and substantially more variable across cities than the corresponding mobility hierarchy, demonstrating that cities with similar mobility organization can differ in how recurrent social overlap is organized across locations. Finally, we show that the spatial proximity of colocation activity to social centers relative to residential centers provides a complementary spatial dimension of urban organization. Together, these results separate three features that are conflated when cities are characterized through mobility alone: movement between locations, local opportunities for social interaction, and the organization of recurrent social overlap across locations.
\end{singlespace}
\end{abstract}

\maketitle

\section{Introduction}

By 2050, nearly 68\% of the global population will reside in urban areas \cite{un2018world}. As cities expand, understanding their urban social fabric becomes increasingly critical for addressing spatial segregation \cite{massey1993american}, inequality \cite{glaeser2009inequality}, and access to opportunity \cite{chetty2022social}. Historically, efforts to quantify urban connectedness relied heavily on static socio-demographic surveys \cite{putnam2000bowling, guest1999social, mcpherson2006social} and qualitative ethnographic observations \cite{jacobs1961death, whyte1980social}. While foundational, these approaches lack the spatial and temporal resolution required to capture the dynamic reality of daily urban interactions. Consequently, planners often rely on the distribution of physical infrastructure itself as a proxy for social integration \cite{klinenberg2018palaces, fraser2024great}. Yet recent shifts in commuting, working, and communication patterns have fundamentally altered how urban populations interact across space \cite{yabe2023behavioral, santana2023covid, barrero2023evolution}, highlighting the need for quantitative frameworks capable of resolving the evolving structure of urban social encounters.

The emergence of computational social science has provided a powerful alternative \cite{lazer2009computational}. Early studies using mobile phone traces established fundamental limits on the predictability of individual human mobility \cite{gonzalez2008understanding,song2010limits,pappalardo2015returners}. Subsequent work showed that predictability depends on recurrent mobility patterns, activity-location contexts, and the social and non-social processes underlying individual trajectories \cite{Panos_2020,Chen_2022,poudyal2024dynamic}.  More recently, attention has shifted from isolated trajectories toward the spatial organization of encounters themselves, recognizing that cities function not merely as containers of movement, but as topological systems that structure opportunities for interaction \cite{Pan_2013, toole2015coupling, samuelsson2021topodiverse}. High-resolution mobility datasets increasingly allow researchers to quantify social overlap, segregation, and homophily through overlapping activity patterns at shared urban locations \cite{moro2021mobility, louail2014mobile}. 

Beyond mobile traces, integrating open-source geographic data enables the precise evaluation of how physical street networks influence equitable access to these shared environments and shape urban scaling \cite{boeing2017osmnx, poudyal2023characterizing, batty2008size}. Similarly, detailed mobility surveys provide the behavioral context necessary to model how mobility-targeted interventions influence complex urban systems \cite{poudyal2025contrasting}. These developments have provided unprecedented insight into how spatial segregation manifests not only through movement itself, but through neighborhood isolation and overlapping patterns of urban social activity \cite{sun2013understanding,nilforoshan2023human,barbosa2021uncovering,wang2018urban}. More broadly, mobility and urban structure have also been shown to shape downstream collective processes, including epidemic spreading and urban welfare \cite{soriano2022modeling,aguilar2022impact,Mimar_2022}.

Yet these outcomes depend not only on where individuals travel, but also on the environments in which populations encounter one another. Although recent work has increasingly resolved patterns of social interaction and segregation from mobility data, cities are still primarily characterized through density, infrastructure, land use, or mobility flows \cite{bettencourt2011bigger, barthelemy2011spatial}. These quantities are essential for understanding how populations distribute and move across urban space \cite{schlapfer2021universal, alessandretti2020scales}, but movement between locations does not uniquely determine the co-presence occurring within them. Two locations may both exhibit high activity, or even be strongly connected through mobility flows, without supporting the same patterns of social interaction \cite{cho2011friendship}. Moreover, substantial co-presence at two locations does not imply that the same populations encounter one another at both. Locations become socially linked when they share co-present populations, reflecting the repeated use of urban spaces and the geographic constraints that shape opportunities for interaction \cite{sekara2016fundamental, onnela2011geographic}. These distinctions separate three features of urban activity: movement between locations, co-presence within locations, and the overlap of co-presence across locations.

To quantify these dimensions, we construct a co-presence overlap network using two complementary measures: colocation and co-connectedness. Colocation quantifies co-presence activity within individual locations, identifying where opportunities for social interaction are concentrated across the city. Co-connectedness measures the number of co-present pairs shared by two locations, identifying environments that participate in overlapping social populations. Importantly, co-connectedness is not a mobility flux or trip count: two locations are connected because they share co-present pairs, rather than because individuals move directly from one to the other. The resulting representation therefore distinguishes local co-presence from the relational structure connecting social environments.

We first examine whether the spatial concentration of colocation differs from that of mobility. Because movement through a location does not uniquely specify the co-presence occurring there, mobility centers need not coincide with centers of social activity. We identify functional centers independently from mobility and colocation and compare their number and spatial arrangement across cities.

We next examine whether the network defined by co-connectedness possesses a hierarchical organization of its own. Previous studies analyzing massive urban transit flows have shown that physical mobility often exhibits a strongly hierarchical organization across location activity levels \cite{bassolas2019hierarchical}. We ask whether recurrent social overlap exhibits a comparable structure. Specifically, we test whether locations preferentially share co-present pairs with locations occupying similar levels of colocation activity. Physical mobility is constrained by sequential movement through geographic space, whereas social overlap can additionally reflect behavioral homophily, recurrent activity patterns, and the repeated use of shared environments \cite{mcpherson2001birds, feld1981focused}. These processes can generate relationships between locations that are not represented by mobility flows alone.

We find that co-connectedness is preferentially concentrated between locations occupying similar or neighboring levels of colocation activity, producing a non-random hierarchical organization of social overlap that that exceeds expectations from randomized activity-level assignments and activity heterogeneity alone~\cite{expert2011uncovering, newman2003mixing, fosdick2018configuring}. We quantify the strength of this organization using the same hierarchy metric previously applied to mobility networks, allowing the two representations to be compared directly. The hierarchy of social overlap is systematically weaker and substantially more variable across cities than the corresponding mobility hierarchy. Cities with similar mobility organization can therefore differ both in where co-presence is concentrated and in how recurrent social overlap is organized across locations.

Finally, social activity has a spatial dimension that is not captured by this hierarchical organization alone. Spatial accessibility, shaped by distance and transportation constraints, remains a central determinant of urban equity and opportunity \cite{handy1997measuring, bertaud2004spatial}. We therefore compare the hierarchical organization of social overlap with the geographic accessibility of socially central locations, distinguishing the network structure of social overlap from the spatial accessibility of social activity. Together, these quantities characterize complementary dimensions of urban social structure that are not captured by mobility flows alone.

\begin{figure}[t!]
    \centering
    \includegraphics[width=\textwidth]{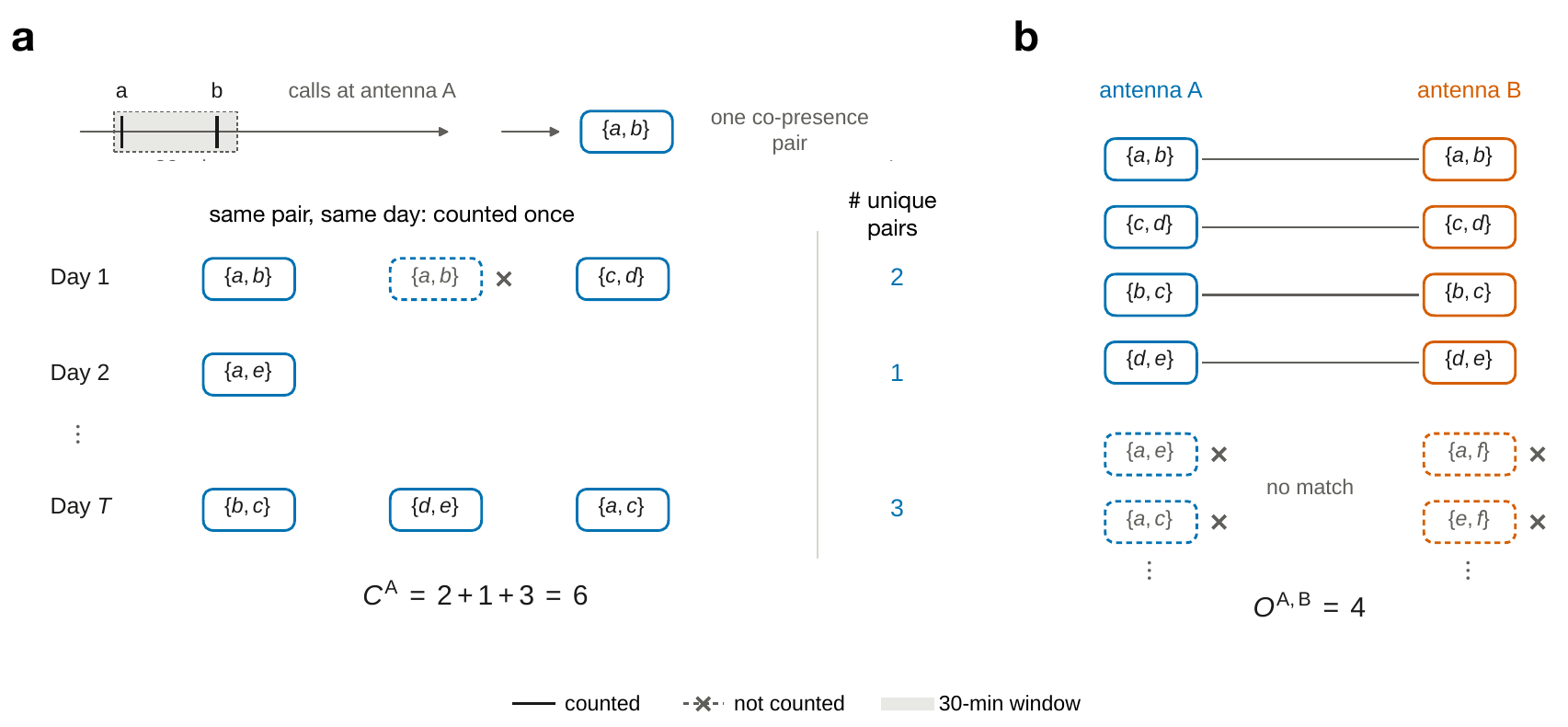}
\caption{\textbf{Definition of the colocation and co-connectedness metrics.} Lower-case letters denote individuals and $\{i,j\}$ a pair of them. \textbf{(a)} Colocation ($C$). Since the receiver's antenna is unknown, a social overlapping event is inferred when a receiver makes a subsequent call from the same antenna as the caller within a 30-minute window. All such events between the same two individuals at the same antenna are counted only once per day; the colocation of an antenna is the resulting number of unique daily pairs, summed over the observation period. \textbf{(b)} Co-connectedness ($O$). Days are pooled first, so each antenna carries the set of pairs co-present there on at least one day, and $O^{AB}$ is the number of pairs common to both sets. Pairs seen at only one antenna do not contribute, and a contributing pair need not have been co-present at the two antennas on the same day.}\label{fig:Fig1}
\end{figure}

\section{Construction of the Co-Presence Overlap Network}

We construct a weighted co-presence overlap network $G=(V,E,W)$, where nodes correspond to urban locations (antennas) and edge weights quantify the number of co-present pairs shared between locations. The construction distinguishes two complementary quantities: colocation, which measures co-presence activity within individual locations, and co-connectedness, which measures the overlap of co-present pairs across locations.

\subsection{Colocation: Local Co-Presence Activity}

We begin by defining a daily co-presence event between individuals $i$ and $j$ at location $A$ on day $d$:

\begin{equation}
E_{ij}^A(d)=
\begin{cases}
1 & \text{if individuals } i \text{ and } j \text{ are co-present at } A \text{ on day } d \\
0 & \text{otherwise}
\end{cases}
\label{eq:event}
\end{equation}

where ``co-present'' requires both individuals to be recorded at location $A$ within a 30-minute time window of each other (we use this threshold throughout the study; see~\ref{fig:figS3} for a robustness analysis). Co-presence is treated as a binary daily event to avoid repeatedly counting the same pair multiple times within a single day. We then define the colocation of location $A$, denoted $C^A$, as the cumulative number of daily co-presence events,

\begin{equation}
C^A = \sum_{i \neq j} \sum_d E_{ij}^A(d)
\label{eq:colocation}
\end{equation}
which quantifies the cumulative co-presence activity observed at location $A$ over the observation period. Higher values of $C^A$ therefore indicate locations at which more co-presence events are observed.

\subsection{Co-connectedness: Overlap Between Social Environments}

To characterize overlap between environments, we define an observation-window indicator,
\begin{equation}
\tilde{E}_{ij}^A =
\begin{cases}
1 & \text{if } \sum_d E_{ij}^A(d) > 0 \\
0 & \text{otherwise.}
\end{cases}
\label{eq:aggregate_event}
\end{equation}
This records whether a pair of individuals was co-present at location $A$ at least once during the entire observation period (March 21--April 19, 2013). It is important to distinguish this observation window from the 30-minute time window used to define individual co-presence events in Eq.~\ref{eq:event}. Using this quantity, we define the co-connectedness between locations $A$ and $B$ as

\begin{equation}
O^{AB} = \sum_{i \neq j} \tilde{E}_{ij}^A \tilde{E}_{ij}^B,
\label{eq:overlap}
\end{equation}
where $O^{AB}$ measures the number of unique pairs of individuals who were co-present at both locations during the observation period. Larger values therefore indicate greater overlap in the co-present pairs associated with the two locations.

\begin{figure}[t!]
    \centering
    \includegraphics[width=.7\textwidth]{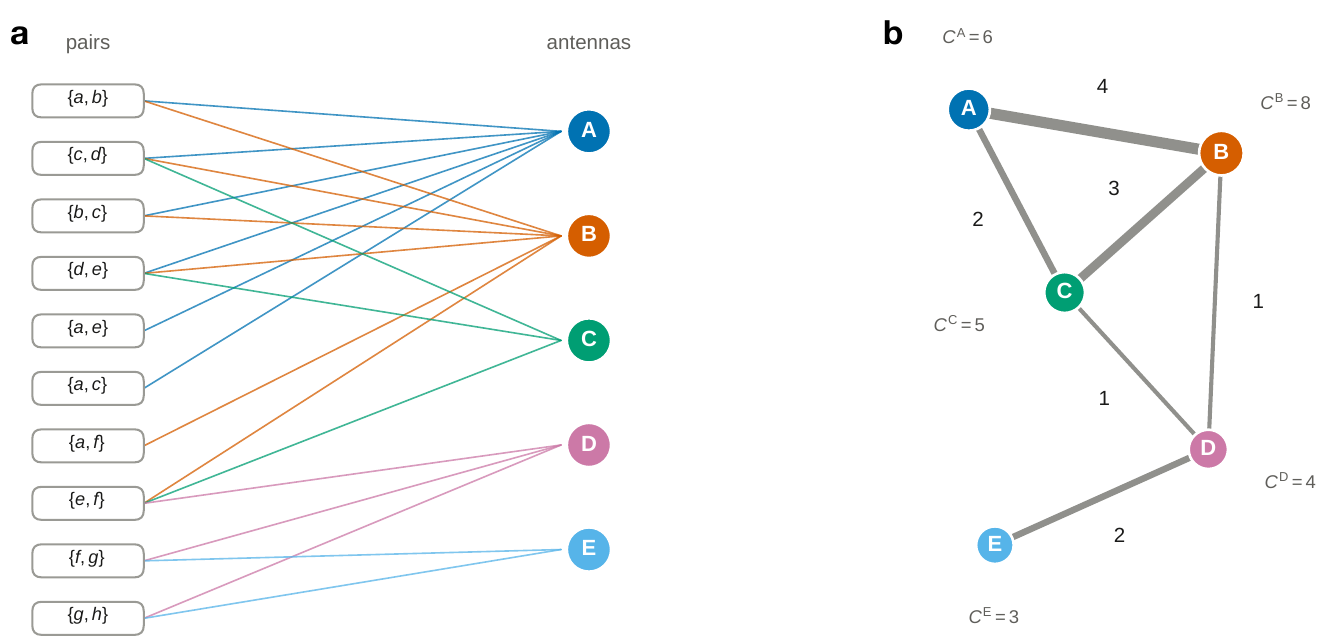}
    \caption{\textbf{Construction of the co-presence overlap network.} Lower-case letters denote individuals and upper-case letters denote antennas. \textbf{(a)} Pair--location incidence. A link joins a pair of individuals to every antenna at which the two were co-present on at least one day (Eq.~\eqref{eq:overlap}); links are colored by antenna. \textbf{(b)} Projection onto locations. Antennas become the nodes of the network, and each pair linked to two antennas contributes one unit to the weight between them, so that $W_{XY}=O^{XY}$ (Eq.~\eqref{eq:overlap}); antennas sharing no pair remain unconnected, so the network is sparse rather than complete. Node size is proportional to the colocation $C^{X}$ (Eq.~\eqref{eq:colocation}). The example reproduces $O^{AB}=4$ from Fig.~\ref{fig:Fig1}.}
    \label{fig:network1}
\end{figure}

The two metrics capture complementary scales of urban organization. Colocation ($C^A$) quantifies cumulative co-presence activity within individual locations, whereas co-connectedness ($O^{AB}$) quantifies the overlap of co-present pairs across locations (Fig.~1). Together, they define a weighted co-presence overlap network in which nodes represent locations and edge weights represent the number of co-present pairs shared between them. This network forms the basis for the hierarchical and spatial analyses that follow. Figure~\ref{fig:network1} illustrates its construction.

\section{Results}

\subsection{Statistical Structure of Urban Social Overlap}

\begin{figure}[t!]
    \centering
    \includegraphics[width=\textwidth]{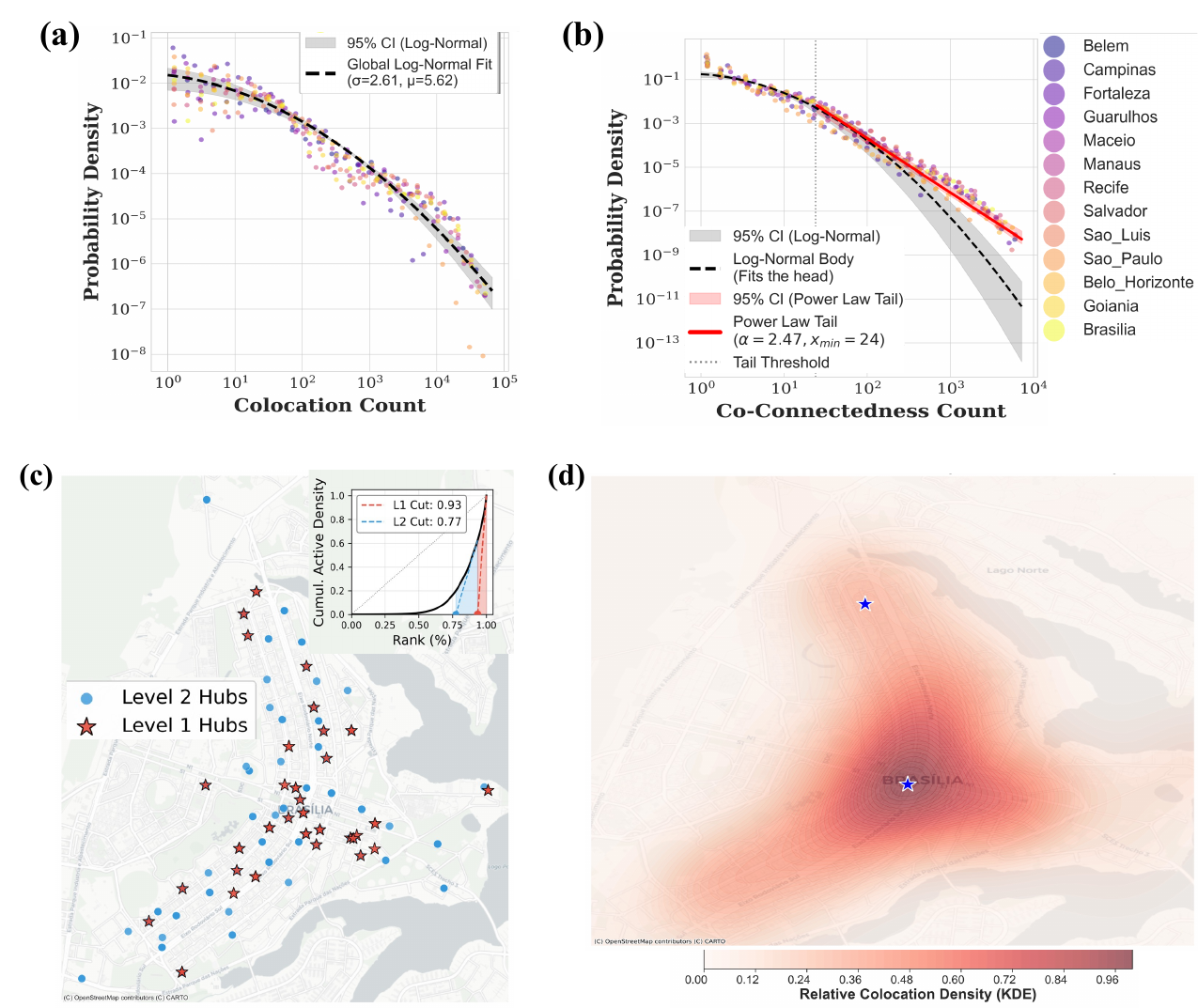}
   \caption{\textbf{Statistical and spatial characteristics of the social overlap network.} \textbf{(a)} Distribution of colocation ($C^A$) across all 13 cities. The dashed line indicates a log-normal fit, consistent with the finite spatial capacity of individual environments. \textbf{(b)} Distribution of co-connectedness ($O^{AB}$).  Weak overlap ties follow a log-normal regime, while highly overlapping location pairs exhibit a scale-free tail. \textbf{(c)} Spatial fragmentation induced by discrete telecommunications infrastructure in Bras\'ilia. Direct thresholding of raw antenna-level colocation volumes identifies numerous adjacent antennas as disconnected hubs (inset). \textbf{(d)} Reconstruction of functional urban centers using kernel density estimation (KDE). Spatial smoothing of antenna-level activity volumes identifies dominant social centers while suppressing infrastructure-level fragmentation. The same procedure is applied consistently across all 13 cities.}
    \label{fig:Fig2}
\end{figure}

To characterize the large-scale structure of urban social overlap, we first examine the statistical distributions of colocation and co-connectedness across 13 Brazilian cities with populations exceeding one million (see~\ref{sec:supp_data} for details of the call detail record dataset and operator coverage). Colocation ($C^A$) consistently follows a log-normal distribution (Fig.~\ref{fig:Fig2}a). This behavior remains robust after normalizing by the geographic coverage area of each antenna (\ref{fig:figS1}), and the correlation between an antenna's Voronoi area and its raw colocation volume is weak in all cities ($r_s \leq 0.29$; \ref{fig:figS2}), despite reaching statistical significance in the largest networks. Together, these results indicate that the observed variation in colocation is not readily explained by differences in the geographic coverage area of the underlying telecommunications infrastructure. The observed log-normal form is consistent with previous evidence that activity within physical environments is constrained by their finite spatial capacity \cite{piazza2025physical,lee2026activity,mitzenmacher2004brief}.

In contrast, co-connectedness ($O^{AB}$) exhibits a qualitatively different distributional structure (Fig.~\ref{fig:Fig2}b). Weak overlap ties follow a similar log-normal regime, whereas highly overlapping location pairs transition into a broad heavy-tailed regime, well described by a power law ($\alpha = 2.47$). This transition indicates that while most location pairs share relatively few co-present pairs, a small number exhibit substantially stronger overlap. This dual-regime structure is robust: sensitivity analyses using tighter ($\Delta t \leq 15$ min) and more relaxed ($\Delta t \leq 45$ min) temporal windows for defining social overlap events confirm that both the log-normal body and the heavy tail persist independently of the chosen threshold (\ref{fig:figS3}).

While these distributions characterize the magnitude of local activity and structural overlap, they do not describe how this organization manifests spatially across the urban landscape. To identify the large-scale backbone of urban social activity, we must detect functional social centers rather than simply cluster discrete telecommunications infrastructure. Because antennas are densely packed within active downtown regions, directly applying LouBar thresholding, which derives activity cutoffs from the curvature of the rank distribution itself \cite{louail2014mobile}, to raw colocation volumes artificially fragments continuous urban cores into dozens of disconnected micro-centers (Fig.~\ref{fig:Fig2}c; see \ref{sec:loubar_levels} for details).  

To recover the underlying spatial organization, we transition from discrete antenna-level measurements to a continuous spatial representation using kernel density estimation (KDE) weighted by raw colocation volumes. This smoothing reconstructs continuous activity landscapes from fragmented infrastructure measurements, allowing dominant social centers to emerge at the scale of the urban environment rather than that of the telecommunications infrastructure. Local maxima of the resulting surface are identified using an adaptive spatial filter whose scale is set by the median separation between active antennas, after which LouBar thresholding is re-applied to isolate the dominant functional centers of each city (Fig.~\ref{fig:Fig2}d). As illustrated for Bras\'ilia, this procedure consolidates fragmented downtown infrastructure into coherent urban cores while filtering out diffuse peripheral activity.

Traditional notions of monocentric and polycentric urban structure are typically inferred from spatial density gradients \cite{clark1951urban}, employment concentration \cite{giuliano1991subcenters}, or commuting flows \cite{roca2009urban, louail2014mobile}. These approaches, central to our understanding of urban morphology, characterize how populations and infrastructure distribute across urban space, but they do not measure how social activity itself organizes spatially. Within our framework, social centers are instead defined by the spatial concentration of colocation activity. They identify regions in which co-presence activity is concentrated, rather than regions that merely concentrate population or movement. The spatial organization of co-presence therefore need not coincide with conventional density- or mobility-based classifications.

\begin{table}[htpb]
\centering
\resizebox{.5\linewidth}{!}{%
\begin{tabular}{l cc @{\hskip 2em} cc}
\toprule
& \multicolumn{2}{c}{\textbf{Mobility}} & \multicolumn{2}{c}{\textbf{Social}} \\
\cmidrule{2-3} \cmidrule{4-5}
\textbf{City} &
\makecell{\textbf{Centers}\\($N_{\mathrm{mob}}$)} &
\textbf{Centricity} &
\makecell{\textbf{Centers}\\($N_{\mathrm{col}}$)} &
\textbf{Centricity} \\
\midrule
S\~ao Paulo    & 2 & Bi-centric  & 5 & Polycentric \\
Fortaleza      & 2 & Bi-centric  & 3 & Polycentric \\
Guarulhos      & 3 & Polycentric & 2 & Bi-centric \\
Campinas       & 1 & Monocentric & 2 & Bi-centric \\
Manaus         & 1 & Monocentric & 2 & Bi-centric \\
Recife         & 1 & Monocentric & 2 & Bi-centric \\
S\~ao Lu\'is   & 1 & Monocentric & 2 & Bi-centric \\
Bras\'ilia     & 2 & Bi-centric  & 2 & Bi-centric \\
Macei\'o       & 2 & Bi-centric  & 2 & Bi-centric \\
Salvador       & 2 & Bi-centric  & 2 & Bi-centric \\
Bel\'em        & 1 & Monocentric & 1 & Monocentric \\
Goi\^ania      & 1 & Monocentric & 1 & Monocentric \\
Belo Horizonte & 1 & Monocentric & 1 & Monocentric \\
\bottomrule
\end{tabular}%
}
\caption{\textbf{Comparison of mobility and social centricity across 13 Brazilian cities.} Mobility and social centers were identified using the same KDE and LouBar thresholding procedure. While several cities exhibit similar overall levels of centricity, substantial differences emerge in the number and location of dominant centers.}
\label{tab:city_centricity}
\end{table}

To evaluate this distinction, we compared social centers with centers identified from mobility flows using the identical KDE and LouBar thresholding procedure (Table~\ref{tab:city_centricity}; see ~\ref{sec:center_detection}). While several cities exhibit similar overall levels of centricity across both representations, important differences emerge. In S\~ao Paulo, mobility identifies two dominant centers, whereas social overlap reveals five. Campinas, Manaus, Recife, and S\~ao Lu\'is appear monocentric from the perspective of mobility but exhibit bi-centric social organization; conversely, Guarulhos appears polycentric under mobility but bi-centric under social overlap. Even when the number of centers is preserved, their locations need not coincide: Bras\'ilia exhibits two dominant centers under both representations, yet the social and mobility hubs occupy substantially different locations (\ref{fig:FigS9}), and similar spatial mismatches are observed in S\~ao Paulo despite the much larger difference in centricity (\ref{fig:FigS10}).

These results indicate that mobility and colocation capture distinct aspects of urban organization: mobility identifies where movement is concentrated, whereas colocation identifies where co-presence activity is concentrated. Cities that appear similar from the perspective of mobility can therefore exhibit different spatial organizations of co-presence. Centricity alone, however, does not describe how locations are related through the populations that use them: cities with similar numbers of social centers may nevertheless differ in how co-present pairs are shared across locations (\ref{fig:FigS6}). We therefore next examine the hierarchical organization of the co-connectedness network.

\subsection{Hierarchical Organization of the Co-presence Overlap Network}

Identifying functional social centers reveals where co-presence activity concentrates, but not how locations are related through recurrent social overlap. We therefore ask whether the co-connectedness network is hierarchically organized, in the sense that locations preferentially share co-present pairs with locations of similar activity, and whether this organization mirrors the hierarchy observed in urban mobility flows. Because mobility hierarchy has been shown to be strongly conserved across cities \cite{bassolas2019hierarchical}, comparing the two reveals whether similar organizations of movement are accompanied by similar organizations of recurrent social overlap.

To measure this structure, we group locations into activity levels using an iterated LouBar procedure applied to the empirical distribution of colocation, ranging from the most active locations to lower-activity locations. This procedure allows the total number of levels ($L$) to emerge dynamically from each city's empirical distribution (\ref{sec:loubar_levels}). We then aggregate co-connectedness across pairs of levels to construct a normalized interaction matrix $T_{ij}$, where $T_{ij}$ represents the fraction of total co-connectedness occurring between locations in levels $i$ and $j$. If overlap is hierarchically organized, the matrix should concentrate near the diagonal, indicating that locations preferentially share co-present pairs with locations in the same or neighboring activity levels. In contrast, a flatter organization would produce substantial overlap between more distant levels.

We summarize this organization using a hierarchy metric, $\Phi$, defined as the fraction of total social overlap contained within the diagonal and immediately adjacent bands of $T_{ij}$:
\begin{equation}
\Phi = \sum_{i,j=1}^{L} T_{ij} \left(\delta_{ij} + \delta_{i(j-1)} + \delta_{(i-1)j}\right),
\label{eq:hierarchy}
\end{equation}
where $L$ is the number of activity levels and $\delta$ is the Kronecker delta. By construction, $\Phi \in (0,1]$, where larger values indicate that co-connectedness is concentrated between locations in the same or adjacent activity levels, whereas smaller values indicate that overlap is distributed more broadly across activity levels.

\begin{figure}[t!]
    \centering
    \includegraphics[width=0.8\textwidth]{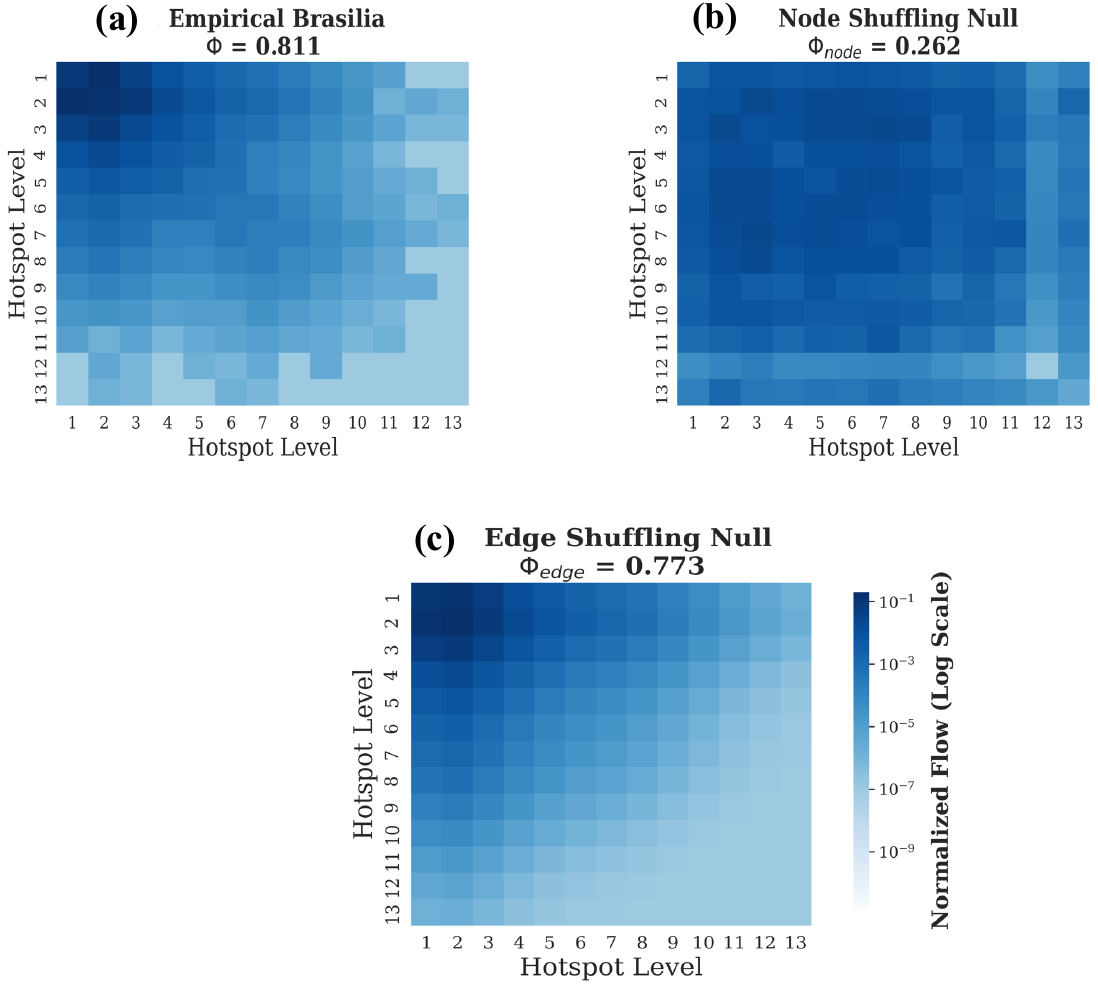}
 \caption{\textbf{Hierarchical organization of social overlap.}
\textbf{(a)} Interaction matrix $T_{ij}$ for Bras\'ilia. Co-connectedness is concentrated near the diagonal, indicating that locations preferentially share co-present pairs with locations in the same or neighboring activity levels ($\Phi=0.811$).
\textbf{(b)} Corresponding Node Shuffling null model. This model preserves the underlying co-connectedness network but randomizes the assignment of activity levels. Removing the correspondence between activity levels and network structure disperses overlap across the matrix, substantially reducing the hierarchy ($\Phi=0.262$, $Z \approx 43$).
\textbf{(c)} Corresponding Edge Shuffling null model. This baseline retains the heterogeneous distribution of overlap across activity levels while randomizing the specific organization of co-connectedness between levels. The Edge Shuffling null yields $\Phi=0.773$, below the empirical value of $\Phi=0.811$, indicating that activity heterogeneity alone does not reproduce the observed degree of hierarchical organization.}  \label{fig:Fig3}
\end{figure}

To illustrate this organization, Fig.~\ref{fig:Fig3}a shows the interaction matrix for Bras\'ilia. Rather than being distributed uniformly across activity levels, co-connectedness is strongly concentrated near the diagonal: locations preferentially share co-present pairs with locations of similar activity, with progressively weaker overlap between more distant levels. This concentration near the diagonal is the defining signature of hierarchical organization in our framework. For Bras\'ilia, the resulting hierarchy is high ($\Phi=0.811$). This value does not imply that overlap is confined within activity levels; rather, it indicates that most co-connectedness occurs within the same level or between adjacent levels.

A high value of $\Phi$ alone does not necessarily imply hierarchical organization. Some diagonal concentration may arise from heterogeneity in colocation activity and overlap volumes rather than from the specific organization of co-connectedness between levels. To distinguish the observed hierarchy from these baseline effects, we compare the empirical networks against two complementary null models (Supplementary Note~\ref{section:null_model}). The first, \emph{Node Shuffling}, randomizes the assignment of activity levels while preserving the empirical co-connectedness network. The second, \emph{Edge Shuffling}, retains the heterogeneous distribution of overlap across activity levels while randomizing the specific organization of co-connectedness between levels. This provides a more stringent baseline in which overlap is driven solely by activity heterogeneity rather than structured organization.
 
In Bras\'ilia, the \emph{Node Shuffling} randomization disperses overlap throughout the interaction matrix, producing a substantially weaker concentration near the diagonal (Fig.~\ref{fig:Fig3}b). The resulting hierarchy decreases from $\Phi_{\mathrm{empirical}}=0.811$ to $\Phi_{\mathrm{node-null}}=0.262$, indicating that the observed diagonal concentration depends strongly on the correspondence between location activity levels and the structure of the co-connectedness network. To determine whether this concentration can arise from activity heterogeneity alone, we also tested the more stringent \emph{Edge Shuffling} model. Across all 13 cities, the empirical hierarchy values also consistently exceed those obtained under Edge Shuffling (\ref{fig:FigS5}), confirming that the observed hierarchical organization cannot be explained by activity heterogeneity alone.
Having established that hierarchical organization is a robust feature of the co-presence overlap network, we next ask whether this hierarchy mirrors the organization of physical mobility flows.

\subsection{Divergence Between Mobility and Social Hierarchies}

\begin{figure}[t!]
    \centering
    \includegraphics[width=\textwidth]{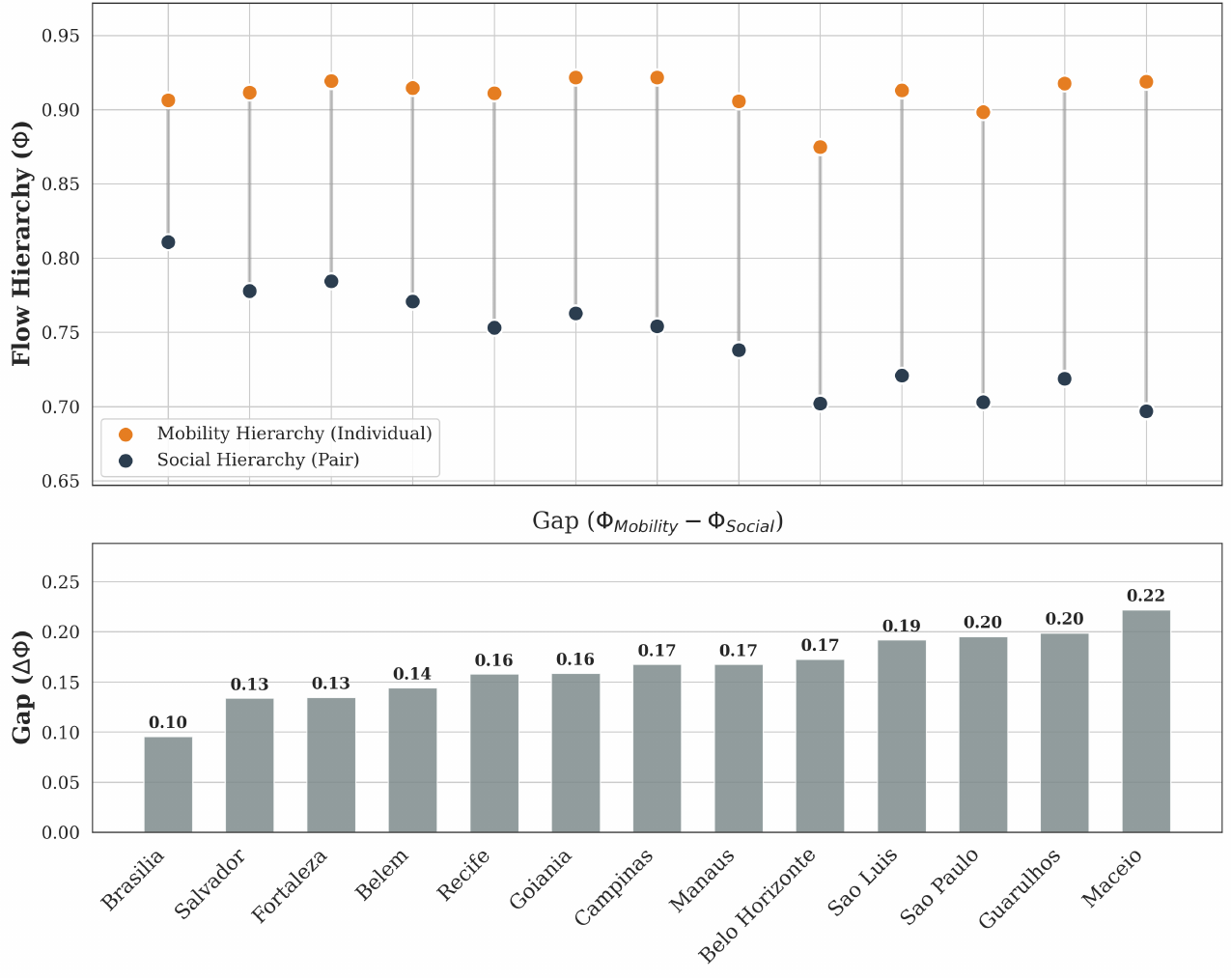}
\caption{\textbf{Mobility and social hierarchies across Brazilian cities.}
\textbf{(Top)} Hierarchy metric ($\Phi$) for mobility networks (orange) and co-presence overlap networks (slate) across the 13 Brazilian cities. Mobility hierarchy varies little across cities, whereas the hierarchy of social overlap exhibits substantially greater variation.
\textbf{(Bottom)} Hierarchy gap, $\Delta\Phi = \Phi_{\mathrm{mobility}} - \Phi_{\mathrm{social}}$, for each city. Positive values indicate that co-connectedness is less concentrated within the same or neighboring activity levels than mobility flows in the corresponding city.}
\label{fig:fig4}
\end{figure}

Consistent with the conserved hierarchical organization previously reported for urban transit flows \cite{bassolas2019hierarchical}, mobility networks exhibit uniformly high hierarchy, with $\Phi_{\mathrm{mobility}}$ clustered around 0.91 across all cities (Fig.~\ref{fig:fig4}). Despite substantial variation in city size, density, and urban form, mobility flows remain concentrated between locations occupying the same or neighboring activity levels. The hierarchical organization of social overlap, in contrast, varies substantially across cities, from $\Phi_{\mathrm{social}}=0.697$ in Macei'o to $\Phi_{\mathrm{social}}=0.811$ in Bras'ilia. Thus, although both networks exhibit hierarchical organization, social overlap is consistently less hierarchical and considerably more variable across cities than mobility.

To quantify this difference, we compute the hierarchy gap,
\begin{equation}
\Delta\Phi = \Phi_{\mathrm{mobility}} - \Phi_{\mathrm{social}},
\label{eq:hierarchy_difference}
\end{equation}
shown in the lower panel of Fig.~\ref{fig:fig4}. Across all cities, $\Delta\Phi$ is positive, ranging from 0.10 in Bras'ilia to 0.22 in Macei'o. Thus, co-connectedness is consistently less concentrated within the same or neighboring activity levels than mobility flows in the corresponding cities. Moreover, cities with nearly identical mobility hierarchies exhibit substantially different social hierarchies, showing that the hierarchical organization of social overlap cannot be inferred from mobility hierarchy alone.

The hierarchy gap also varies with mobility centricity in this sample. The seven cities that are monocentric with respect to mobility (Table~\ref{tab:city_centricity}) occupy an intermediate range of $\Delta\Phi$ ($0.14$--$0.19$), whereas cities with multiple mobility centers occur at both lower and higher values ($\Delta\Phi \leq 0.13$ or $\Delta\Phi \geq 0.20$). Because $\Phi_{\mathrm{mobility}}$ varies little across cities, these differences primarily reflect variation in the hierarchical organization of social overlap. Thus, while monocentric cities in our sample exhibit a relatively narrow range of social hierarchy, multicentric cities span a broader range, from Bras'ilia at one extreme to Macei'o and S\~ao Paulo at the other.

\subsection{Catchment Efficiency and Hierarchy of Social Overlap}

The results thus far indicate that the hierarchical organization of social overlap varies substantially across cities. We next examine how this variation relates to the spatial distribution of social activity. While the hierarchy metric ($\Phi$) quantifies how co-connectedness is organized across activity levels, it does not capture how social activity is distributed spatially relative to the city's functional and residential centers.

\begin{table}[h!]
    \centering
        \setlength{\tabcolsep}{8pt}
    \resizebox{0.7\linewidth}{!}{%
    \begin{tabular}{lcccc}
        \toprule
        \textbf{City} &
        \makecell{\textbf{Res. Centers}\\($N_{\mathrm{res}}$)} &
        \makecell{\textbf{Coloc. Centers}\\($N_{\mathrm{col}}$)} &
        \makecell{\textbf{Catchment Efficiency}\\($\eta$)} &
        \makecell{\textbf{Social Hierarchy}\\($\Phi$)} \\
        \midrule
        Bras\'ilia     & 2 & 2 & $+51.69\%$ & 0.811 \\
        S\~ao Paulo    & 4 & 5 & $+48.43\%$ & 0.703 \\
        Fortaleza      & 2 & 3 & $+41.87\%$ & 0.784 \\
        Campinas       & 2 & 2 & $+32.97\%$ & 0.754 \\
        Guarulhos      & 2 & 2 & $+32.46\%$ & 0.719 \\
        S\~ao Lu\'is   & 2 & 2 & $+24.53\%$ & 0.721 \\
        Macei\'o       & 2 & 2 & $+8.55\%$  & 0.697 \\
        Recife         & 4 & 2 & $-0.47\%$  & 0.753 \\
        Salvador       & 4 & 2 & $-20.42\%$ & 0.778 \\
        Manaus         & 4 & 2 & $-24.66\%$ & 0.738 \\
        Goi\^ania      & 4 & 1 & $-24.92\%$ & 0.763 \\
        Belo Horizonte & 3 & 1 & $-32.67\%$ & 0.702 \\
        Bel\'em        & 2 & 1 & $-35.16\%$ & 0.771 \\
        \bottomrule
    \end{tabular}%
    }
   \caption{\textbf{Catchment efficiency and social hierarchy across Brazilian cities.} Catchment efficiency ($\eta$) quantifies the relative proximity of colocation activity to social centers compared with residential centers, while social hierarchy ($\Phi$) quantifies the hierarchical organization of social overlap. Cities are sorted by catchment efficiency.}    \label{tab:catchment_efficiency}
\end{table}

To characterize this spatial dimension, we compare the proximity of colocation activity to functional social centers and to residential centers identified from high-resolution population data (see~\ref{sec:residential_centers}). Let $\mathcal{N}_{\mathrm{col}}$ denote the set of social centers and $\mathcal{N}_{\mathrm{res}}$ the set of residential centers. For each location $i$ with colocation volume $w_i$, we compute the colocation-weighted distance to the nearest residential center,
\begin{equation}
D_{\mathrm{res}}=\sum_i w_i \min_{n\in\mathcal{N}_{\mathrm{res}}} d(i,n),
\end{equation}
and the colocation-weighted distance to the nearest social center,
\begin{equation}
D_{\mathrm{col}}=\sum_i w_i \min_{n\in\mathcal{N}_{\mathrm{col}}} d(i,n).
\end{equation}
We then define the catchment efficiency,
\begin{equation}
\eta=\left(\frac{D_{\mathrm{res}}-D_{\mathrm{col}}}{D_{\mathrm{res}}}\right)\times100\%,
\end{equation}
which measures the relative proximity of colocation activity to social centers compared with residential centers. Positive values indicate that colocation activity lies, on average, closer to a social center than to a residential center, whereas negative values indicate the opposite.
Together, catchment efficiency ($\eta$) and social hierarchy ($\Phi$) capture complementary dimensions of urban social organization. Whereas $\Phi$ characterizes the hierarchical organization of social overlap, $\eta$ characterizes the spatial proximity of colocation activity to social centers relative to residential centers.

\begin{figure}[t!]
    \centering
    \begin{adjustbox}{center}
    \includegraphics[width=\textwidth]{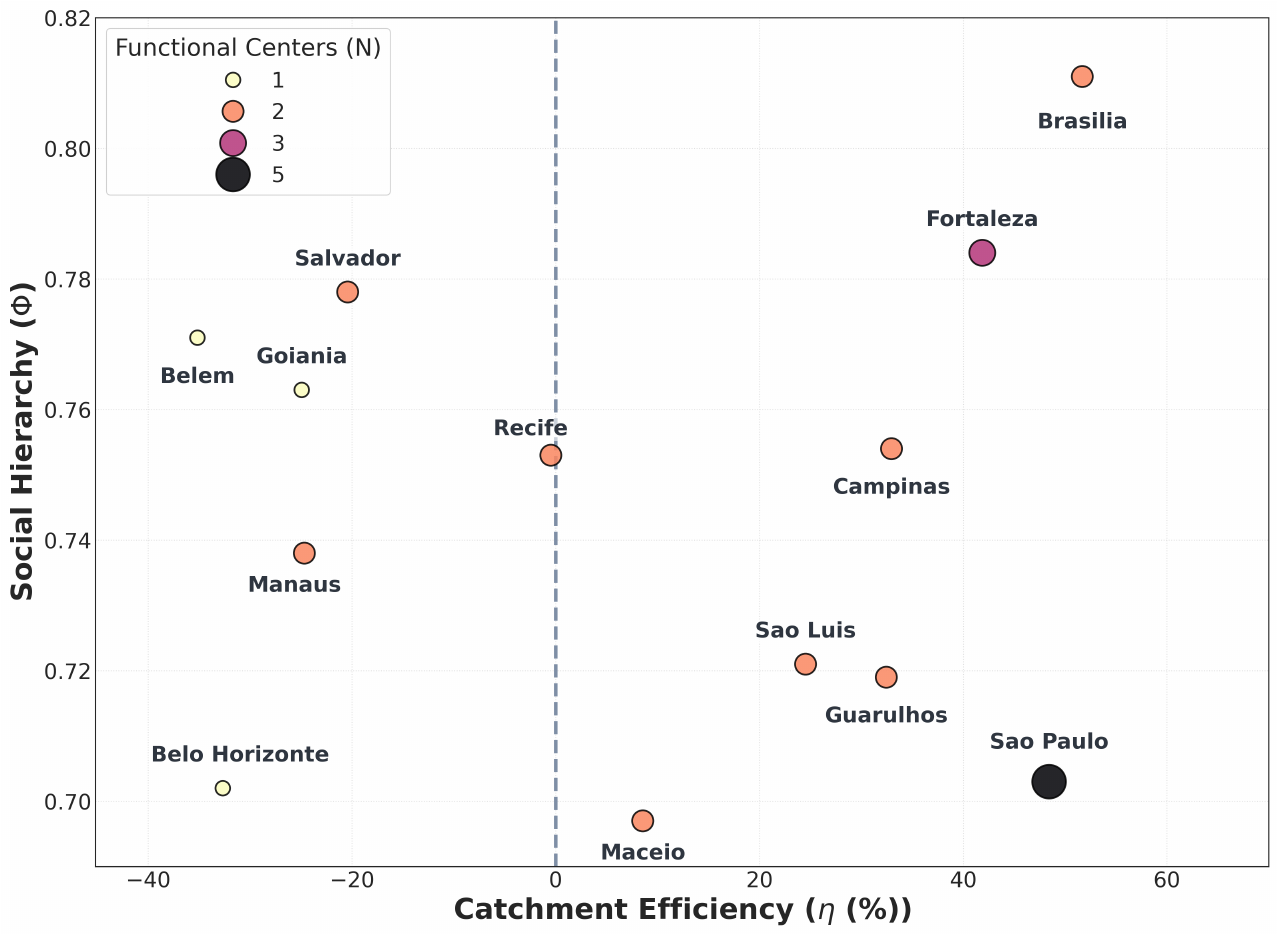}
    \end{adjustbox}
\caption{\textbf{Catchment efficiency and hierarchy of social overlap.} Each point represents a city positioned according to its catchment efficiency ($\eta$) and social hierarchy ($\Phi$). Cities occupy a broad range of positions along both dimensions. The dashed vertical line ($\eta=0$) separates cities in which colocation activity lies, on average, closer to a social center than to a residential center ($\eta>0$) from those in which it does not. Bubble size indicates the number of functional social centers.}  
\label{fig:Fig5}
\end{figure}

Table~\ref{tab:catchment_efficiency} summarizes both metrics across the 13 cities, revealing substantial variation in catchment efficiency, from $\eta=-35.16\%$ in Bel\'em to $\eta=+51.69\%$ in Bras\'ilia. Figure~\ref{fig:Fig5} combines catchment efficiency with social hierarchy to position each city along these two dimensions. Bras\'ilia combines the highest observed social hierarchy ($\Phi=0.811$) with the highest catchment efficiency ($\eta=+51.69\%$), whereas S\~ao Paulo exhibits similarly high catchment efficiency ($\eta=+48.43\%$) but one of the weakest social hierarchies ($\Phi=0.703$). Other cities occupy intermediate positions between these extremes. Together, these results show that cities with similar spatial relationships between colocation activity and social centers can exhibit markedly different hierarchical organization of social overlap, while cities with similar hierarchy can differ substantially in their catchment efficiency. The two measures therefore capture complementary dimensions of urban social organization.

\section{Discussion}

Cities are often characterized through the distribution of population, infrastructure, and mobility flows~\cite{Lee2017, Barbosa2018}. While these measures describe where people live and how they move, they do not uniquely determine the co-presence that occurs within locations or the extent to which locations share the same co-present pairs. In this work, we introduced a co-presence overlap framework that separates these dimensions of urban activity. Colocation quantifies co-presence activity within individual locations, while co-connectedness captures recurrent social overlap across locations. Across 13 Brazilian cities, we find that both the spatial concentration of colocation and the hierarchical organization of co-connectedness contain information that is not captured by mobility flows alone.

The distinction is already apparent in the spatial concentration of colocation. Although social and mobility centers are derived from the same underlying data and identified using the same methodological procedure, their number and location often differ substantially. Several cities that appear monocentric from the perspective of mobility exhibit multiple centers of colocation activity, while others display similar levels of centricity but different spatial arrangements of their dominant centers. Mobility therefore identifies where movement is concentrated, whereas colocation identifies where co-presence activity is concentrated. The two need not coincide, demonstrating that the geography of co-presence cannot be inferred directly from the geography of movement.

Co-connectedness reveals a second level of organization by identifying locations that share the same co-present pairs. Across all cities, co-connectedness is preferentially concentrated between locations occupying similar or neighboring levels of colocation activity, producing a hierarchical organization of social overlap. Comparisons with randomized null models show that this concentration cannot be reproduced by activity heterogeneity alone. The result therefore goes beyond the spatial concentration of colocation: locations are related not simply according to how much co-presence activity they contain, but according to which co-present pairs they share and how those relationships are distributed across activity levels. This provides a relational description of urban social organization that is absent from local activity measures alone.

A central result of this study is that the strength of this hierarchical organization differs markedly from that observed in mobility flows. Previous work has shown that urban mobility networks exhibit a strongly conserved hierarchy across cities~\cite{bassolas2019hierarchical}, and we recover this pattern in all 13 cities analyzed here. Social overlap is also hierarchical, but its hierarchy is consistently weaker and substantially more variable. Cities with nearly identical mobility hierarchies can therefore exhibit markedly different organizations of recurrent social overlap. The consistency of mobility hierarchy may reflect common constraints imposed by transportation systems and the need to traverse urban space efficiently~\cite{Kirkley_2018,gallotti2014anatomy}. Co-connectedness, by contrast, can additionally reflect recurrent patterns of activity, the repeated use of shared environments, and the social processes that determine which individuals encounter one another. While our analysis does not identify the mechanisms responsible for the observed variation, it demonstrates that the hierarchical organization of recurrent social overlap cannot be inferred from mobility hierarchy alone.

The comparison across cities further suggests that this divergence is related to urban centricity, although the limited number of cities warrants caution in interpreting this association. Cities that are monocentric with respect to mobility occupy a relatively narrow intermediate range of the hierarchy gap, whereas cities with multiple mobility centers span both extremes. Because mobility hierarchy itself varies little across cities, this pattern primarily reflects variation in the hierarchy of social overlap. Multicentric cities in our sample can therefore exhibit either relatively similar mobility and social hierarchies, as in Bras\'ilia, or much larger differences between them, as in Macei\'o and S\~ao Paulo. Whether urban centricity systematically shapes the organization of recurrent social overlap, or whether both arise from other features of urban form and behavior, remains an open question.

The spatial organization of colocation provides a complementary dimension. We introduced catchment efficiency ($\eta$) to compare the proximity of colocation activity to functional social centers with its proximity to residential centers. This quantity is distinct from the hierarchy metric: $\Phi$ characterizes how co-connectedness is distributed across location activity levels, whereas $\eta$ characterizes the spatial relationship between colocation activity and the two sets of centers. The two need not covary closely across the cities in our sample; Bras\'ilia and S\~ao Paulo, for example, have similarly high catchment efficiencies but substantially different social hierarchies. Conversely, cities with similar values of $\Phi$ can differ considerably in catchment efficiency. Hierarchy and catchment efficiency therefore describe complementary properties of urban social organization, one concerning the network organization of recurrent overlap and the other the spatial concentration of co-presence relative to functional and residential centers.

More broadly, the framework separates aspects of urban activity that are often combined when cities are represented through mobility alone. Previous work has shown that mobility traces provide powerful insight into human movement~\cite{gonzalez2008understanding}, segregation~\cite{xu2019quantifying}, and urban structure~\cite{louail2014mobile}. Our results show that the same movement system can support different spatial concentrations of co-presence and different organizations of recurrent social overlap. These distinctions may be relevant whenever outcomes depend on who repeatedly shares environments rather than simply on where individuals travel. Co-presence and repeated exposure are central to questions of social integration, access to opportunity, and segregation~\cite{wang2018urban,small2019role}, while contact structure is also an important component of models of epidemic dynamics, outbreak source inference, and targeted public-health interventions~\cite{Hazarie_2021,soriano2022modeling,aguilar2022impact,Ansari_2022,valganon2023}. Establishing how the quantities introduced here relate to these downstream processes will require analyses designed specifically around those outcomes, but the co-presence overlap framework provides a way to distinguish the contribution of local co-presence and recurrent overlap from movement itself. Because these relationships are inherently relational, they contain information that is not directly represented by aggregate flows or static spatial distributions~\cite{sun2013understanding}.

Several limitations should be noted. First, our analysis relies on Call Detail Records (CDRs), which restrict temporal and spatial resolution to the coverage time and areas of telecommunications antennas. Although the KDE-based procedure mitigates fragmentation arising from antenna placement, higher-resolution location data could provide a more detailed characterization of colocation centers and overlap structures. Second, co-presence is inferred indirectly from spatiotemporal coincidence rather than from direct observation of face-to-face interaction or underlying social relationships. The networks constructed here should therefore be interpreted as representations of inferred co-presence and recurrent opportunities for interaction rather than confirmed social encounters. Finally, our analysis focuses on 13 large Brazilian cities and average mobility patterns without accounting for sociodemographic differences~\cite{macedo2022differences,macedo2026parenthood}. Extending this framework to cities with different historical, economic, demographic, and transportation contexts will be important for assessing the generality of the patterns reported here and for identifying the processes that generate variation in urban social overlap.

Taken together, our results separate three dimensions of urban organization that are not equivalent: movement between locations, co-presence activity within locations, and recurrent social overlap across locations. Colocation reveals a spatial organization that can differ from that inferred from mobility, while co-connectedness reveals a hierarchical organization whose strength varies substantially even among cities with similar mobility hierarchies. Catchment efficiency provides an additional spatial characterization of where colocation activity lies relative to functional and residential centers. These results show that movement alone does not determine either where co-presence concentrates or how recurrent social environments are related across the city. A fuller description of urban organization therefore requires considering not only how people move through space, but also where they repeatedly encounter one another and how those shared environments are connected.

\bibliography{ms}

\clearpage

\clearpage
\setcounter{section}{0}
\setcounter{subsection}{0}
\setcounter{subsubsection}{0}
\setcounter{figure}{0}
\setcounter{table}{0}
\setcounter{equation}{0}
\setcounter{page}{1}
\setlength{\parindent}{0pt}

\makeatletter
\titleformat{\section}
  {\normalfont\bfseries\large}
  {}
  {0pt}
  {\ifnum\value{section}>0
     \bfseries\large Supplementary Note \arabic{section}.\ 
   \fi}
\renewcommand*{\thesection}{Supplementary Note \arabic{section}}
\renewcommand*{\thesubsection}{\arabic{section}.\arabic{subsection}}
\renewcommand*{\p@subsection}{}
\renewcommand*{\thesubsubsection}{\thesubsection.\arabic{subsubsection}}
\renewcommand*{\p@subsubsection}{}
\renewcommand{\fnum@figure}{\thefigure}
\makeatother
\renewcommand{\thefigure}{Supplementary Figure \arabic{figure}}
\renewcommand{\thetable}{\arabic{table}}
\renewcommand{\theequation}{S\arabic{equation}}
\renewcommand{\thepage}{\arabic{page}}
\renewcommand{\sectionautorefname}{Section}
\renewcommand{\subsectionautorefname}{Section}
\renewcommand{\subsubsectionautorefname}{Section}

\newcommand{\floor}[1]{\lfloor #1 \rfloor}
\newcommand{\todo}[1]{}
\renewcommand{\todo}[1]{{\color{orange}[[TODO: {#1}]]}}
\newcommand{\jpb}[1]{\textit{\color{blue}[[JPB: {#1}]]}}
\DeclareRobustCommand{\jfcom}[1]{\noindent{\sethlcolor{magentish}\hl{\textbf{GG COMMENT:}  #1}}}
\DeclareRobustCommand{\highlight}[1]{\noindent{\sethlcolor{greenish}\hl{ #1}}}

\begin{flushleft}
{\huge Supplementary Information}\\[0.3cm]
{\large\textbf{The hierarchical organization of urban social overlap: How shared social spaces diverge from physical mobility}}
\end{flushleft}

\begingroup
\setlength{\parindent}{0pt}
\renewcommand{\arraystretch}{1.08}
\begin{tabular*}{\textwidth}{@{\extracolsep{\fill}}lr}
\textbf{Supplementary Note 1. Data} & \pageref{sec:supp_data} \\
\textbf{Supplementary Note 2. Robustness to Antenna Coverage Area} & \pageref{sec:supp_area} \\
\textbf{Supplementary Note 3. Time-Window Analysis} & \pageref{sec:supp_timewindow} \\
\textbf{Supplementary Note 4. Iterated LouBar Procedure and Activity Levels ($L$)} & \pageref{sec:supp_loubar} \\
\textbf{Supplementary Note 5. Null Models} & \pageref{section:null_model} \\
\hspace{1.5em}5.1. Node Shuffling & \pageref{sec:supp_node_shuffle} \\
\hspace{1.5em}5.2. Edge Shuffling & \pageref{sec:supp_edge_shuffle} \\
\hspace{1.5em}5.3. Statistical Evaluation & \pageref{sec:supp_stats} \\
\hspace{1.5em}5.4. Evaluating Structural Robustness: Node and Edge Shuffling & \pageref{sec:supp_robust_null} \\
\textbf{Supplementary Note 6. Difference between Colocation and Mobility Centers} & \pageref{sec:supp_centers} \\
\textbf{Supplementary Note 7. Geographic Embedding and Urban Form} & \pageref{section:geographic_embedding} \\
\textbf{Supplementary Note 8. Spatial Consolidation and Center Detection} & \pageref{sec:center_detection} \\
\hspace{1.5em}8.1. Determining the Spatial Scale & \pageref{sec:supp_spatial_scale} \\
\hspace{1.5em}8.2. Continuous Kernel Density Estimation (KDE) & \pageref{sec:supp_kde} \\
\hspace{1.5em}8.3. Local Maximum Detection and the LouBar Procedure & \pageref{sec:supp_peak_loubar} \\
\hspace{1.5em}8.4. Bandwidth Sensitivity Analysis & \pageref{sec:supp_bandwidth} \\
\textbf{Supplementary Note 9. Residential Population Assignment} & \pageref{sec:residential_centers} \\
\textbf{Supplementary References} & \pageref{sec:supp_refs} \\
\end{tabular*}
\endgroup
\newpage

\section{Data}
\label{sec:supp_data}

The raw dataset was provided for research purposes by a telecommunications operator in Brazil. The identity of the operator and detailed market share information cannot be disclosed for contractual reasons. The raw data consist of 3.1 billion anonymized call detail records (CDRs) collected between March 21 and April 19, 2013, corresponding to a typical 30-day period without major holidays or large-scale events. At the time of data collection, mobile phones were used by approximately 85\% of the Brazilian urban population, and the operator maintained an average market share of roughly 20\% across the analyzed cities, although this varied between municipalities.

The dataset primarily contains outgoing voice calls placed by users on the operator's network and does not include incoming calls, SMS traffic, or internet activity. Both caller and recipient identifiers are available, allowing communication between users belonging to the same operator to be identified. However, only the location of the caller is observed at the time of a call; the precise location of the recipient cannot be determined from the corresponding record. Co-presence is therefore inferred using the temporal procedure described in the main text rather than observed directly. At the time of data collection, voice-over-IP services had not yet reached widespread adoption in Brazil, making voice calls the dominant mode of mobile communication. Each record contains the date, time, and duration of the call, encrypted identifiers for both the caller and recipient, information about the originating and destination operators, and the geographic coordinates of the serving antenna. User locations are inferred from the latitude and longitude of the serving antenna. All records were anonymized prior to analysis.

The resulting social overlap networks span a broad range of urban scales and network sizes. Summary statistics for all 13 cities are provided in Supplementary Table~\ref{tab:dataset_stats}. These 13 cities are the largest cities available in our dataset with populations exceeding one million inhabitants.

\begin{table}[htpb]
\centering
\resizebox{\textwidth}{!}{%
\begin{tabular}{lrrrrrrrr}
\toprule
\textbf{City} & \textbf{Nodes ($N$)} & \textbf{Edges ($E$)} & \textbf{Total $C^A$} & \textbf{Total $O^{AB}$} & \textbf{$\max C^A$} & \textbf{$med. C^A$} & \textbf{$\max O^{AB}$} & \textbf{$med. O^{AB}$} \\
\midrule
Bel\'em & 680 & 87,118 & 2,934,632 & 1,016,479 & 48,493 & 1292.00 & 7,202 & 3.00 \\
Campinas & 504 & 14,500 & 150,027 & 68,994 & 7,652 & 38.00 & 1,245 & 2.00 \\
Fortaleza & 1,196 & 302,883 & 8,041,030 & 6,113,424 & 65,366 & 2327.00 & 6,128 & 10.00 \\
Guarulhos & 353 & 14,915 & 628,983 & 233,728 & 25,977 & 218.50 & 2,630 & 4.00 \\
Macei\'o & 420 & 36,115 & 1,604,667 & 983,086 & 65,528 & 734.50 & 4,258 & 9.00 \\
Manaus & 666 & 52,983 & 1,894,282 & 632,833 & 50,069 & 197.00 & 7,212 & 2.00 \\
Recife & 713 & 99,000 & 3,169,924 & 1,864,184 & 40,367 & 1760.00 & 4,756 & 6.00 \\
Salvador & 1,442 & 288,067 & 6,887,370 & 4,131,097 & 54,293 & 686.00 & 6,629 & 6.00 \\
S\~ao Lu\'is & 547 & 74,678 & 2,369,659 & 1,824,876 & 37,011 & 2108.00 & 4,351 & 6.00 \\
S\~ao Paulo & 4,792 & 323,150 & 2,937,634 & 1,436,061 & 61,139 & 64.00 & 2,005 & 1.00 \\
Belo Horizonte & 1,604 & 276,545 & 4,004,197 & 1,654,470 & 50,566 & 389.00 & 4,653 & 2.00 \\
Goi\^ania & 913 & 109,808 & 4,725,327 & 1,906,211 & 65,596 & 1106.50 & 6,718 & 7.00 \\
Bras\'ilia & 730 & 56,985 & 3,259,631 & 924,603 & 66,617 & 304.00 & 5,489 & 4.00 \\
\bottomrule
\end{tabular}%
}
\caption{\textbf{Summary statistics of the social overlap networks.} Network size (nodes $N$ and edges $E$), colocation ($C^A$), and co-connectedness ($O^{AB}$) statistics for the 13 cities analyzed in this study. Because the distributions of $C^A$ and $O^{AB}$ are highly skewed, we report the median rather than the mean as a measure of central tendency.}
\label{tab:dataset_stats}
\end{table}
\newpage
\section{Robustness to Antenna Coverage Area}\label{sec:supp_area}

Given that antenna coverage areas vary substantially across cities, apparent differences in colocation could partly reflect variation in the spatial resolution of the telecommunications infrastructure. To assess this possibility, we examine colocation normalized by antenna Voronoi area and quantify the relationship between antenna coverage area and colocation volume. Together, these analyses indicate that the observed heterogeneity in colocation is not readily explained by differences in antenna coverage area.

\begin{figure}[htpb]
    \centering
    \includegraphics[width=\textwidth]{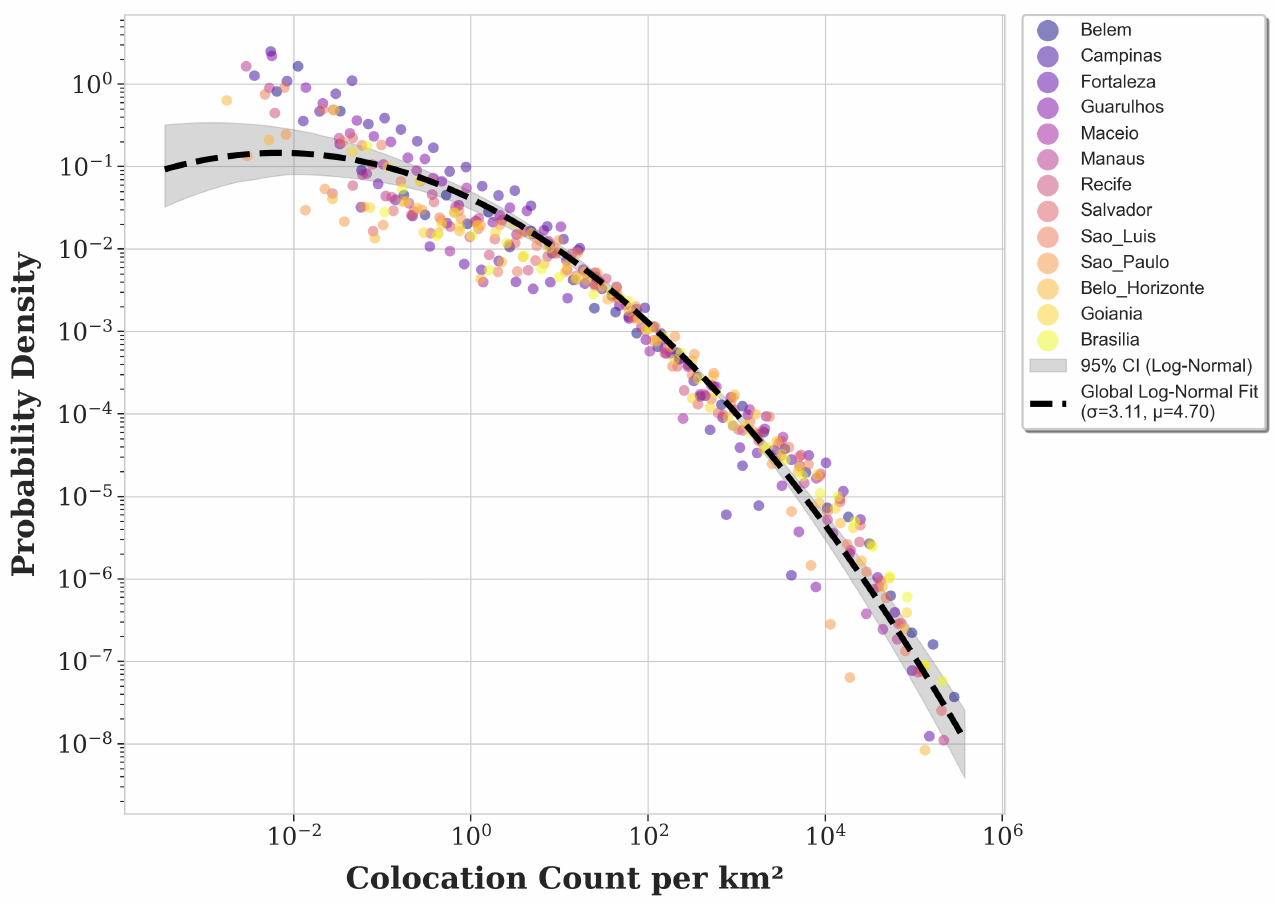}
    \caption{\textbf{Area-normalized colocation distributions.} Probability density distributions of colocation normalized by the Voronoi area of each antenna. The distributions remain broadly log-normal after accounting for differences in antenna coverage area, indicating that the observed statistical form of colocation is robust to variation in spatial resolution.}
    \label{fig:figS1}
\end{figure}

\begin{figure}[htpb]
    \centering
    \includegraphics[width=\textwidth]{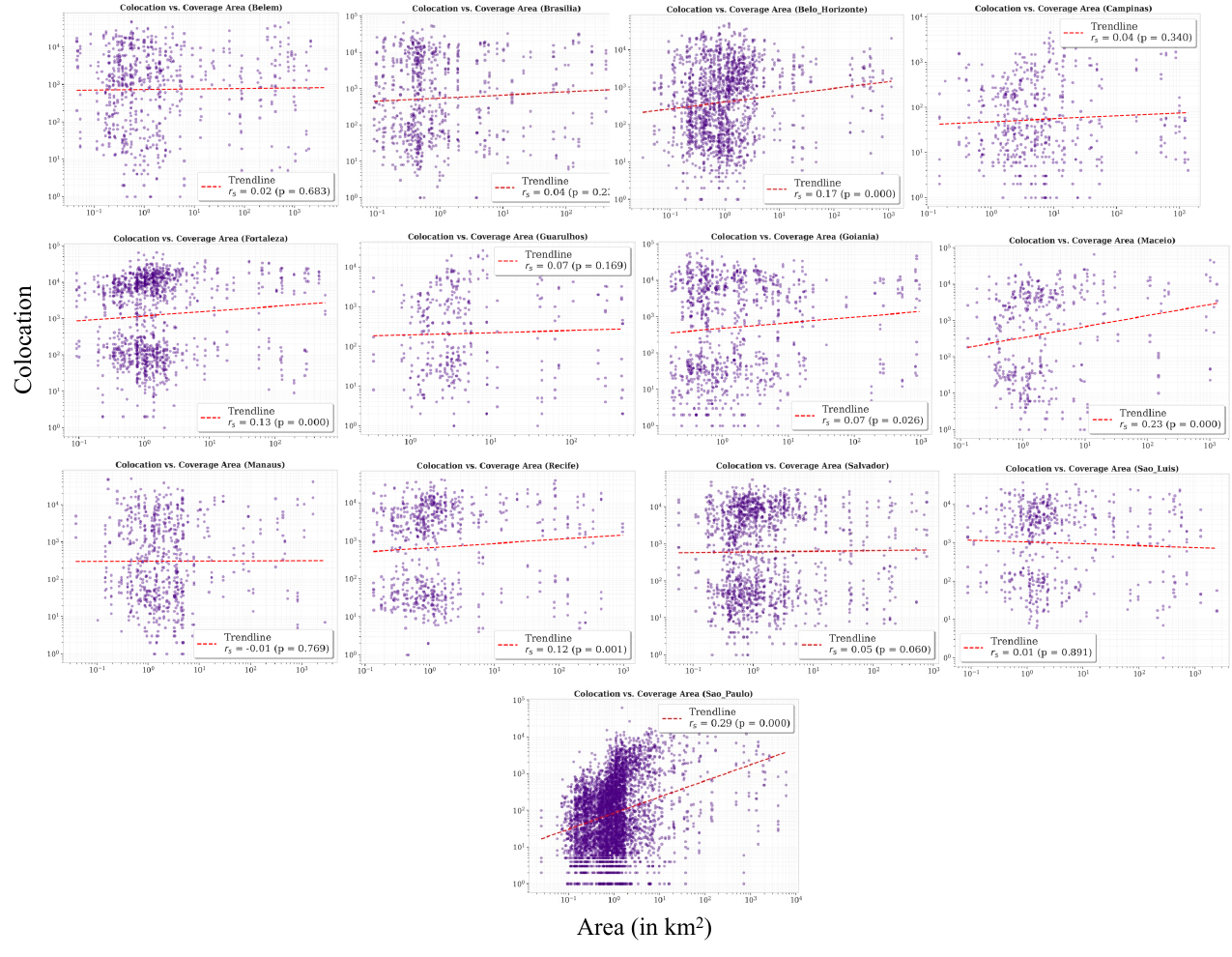}
    \caption{\textbf{Relationship between antenna coverage area and colocation volume.} Scatter plots of antenna Voronoi area and colocation volume across the 13 cities. Correlations are weak in all cases, indicating that variation in colocation volume is not strongly associated with antenna coverage area.}
    \label{fig:figS2}
\end{figure}

\section{Time-Window Analysis}\label{sec:supp_timewindow}

To evaluate the robustness of our results to the temporal definition of co-presence, we recompute the distributions of colocation and co-connectedness using stricter ($\Delta t \leq 15$ min) and more relaxed ($\Delta t \leq 45$ min) temporal thresholds, in addition to the 30-minute threshold adopted throughout the main text.

\begin{figure}[htpb]
    \centering
    \includegraphics[width=\textwidth]{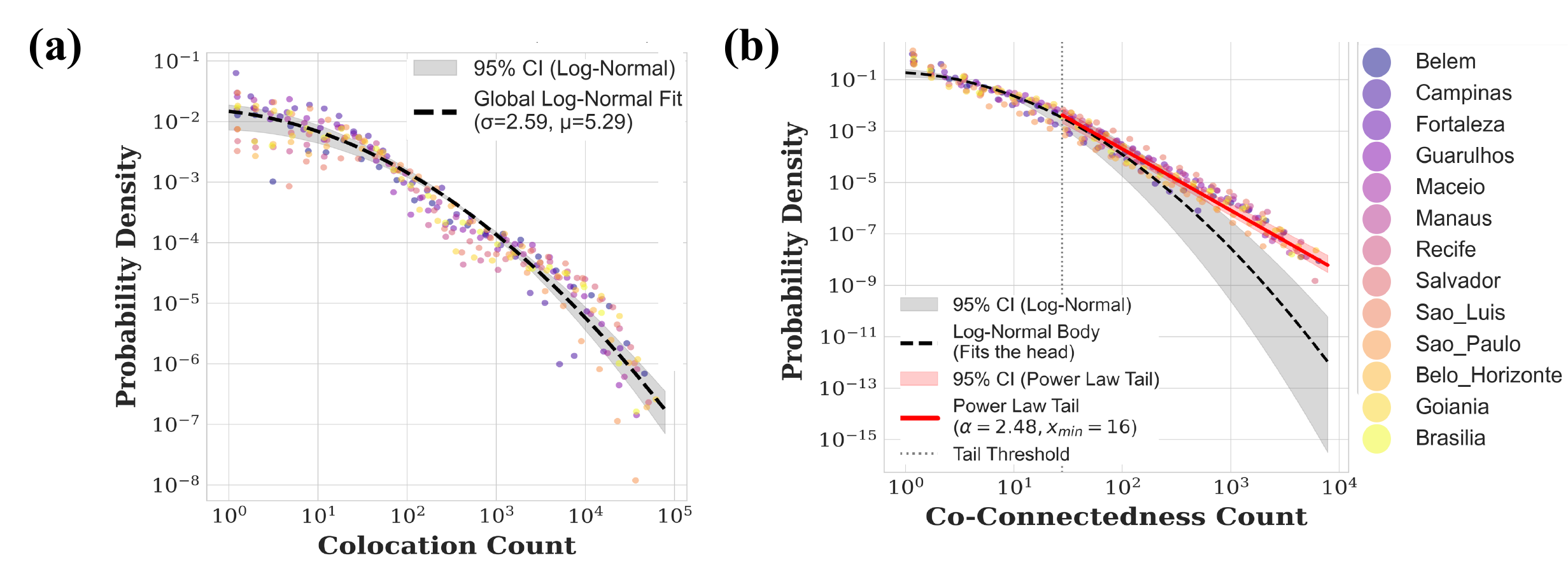}
    \includegraphics[width=\textwidth]{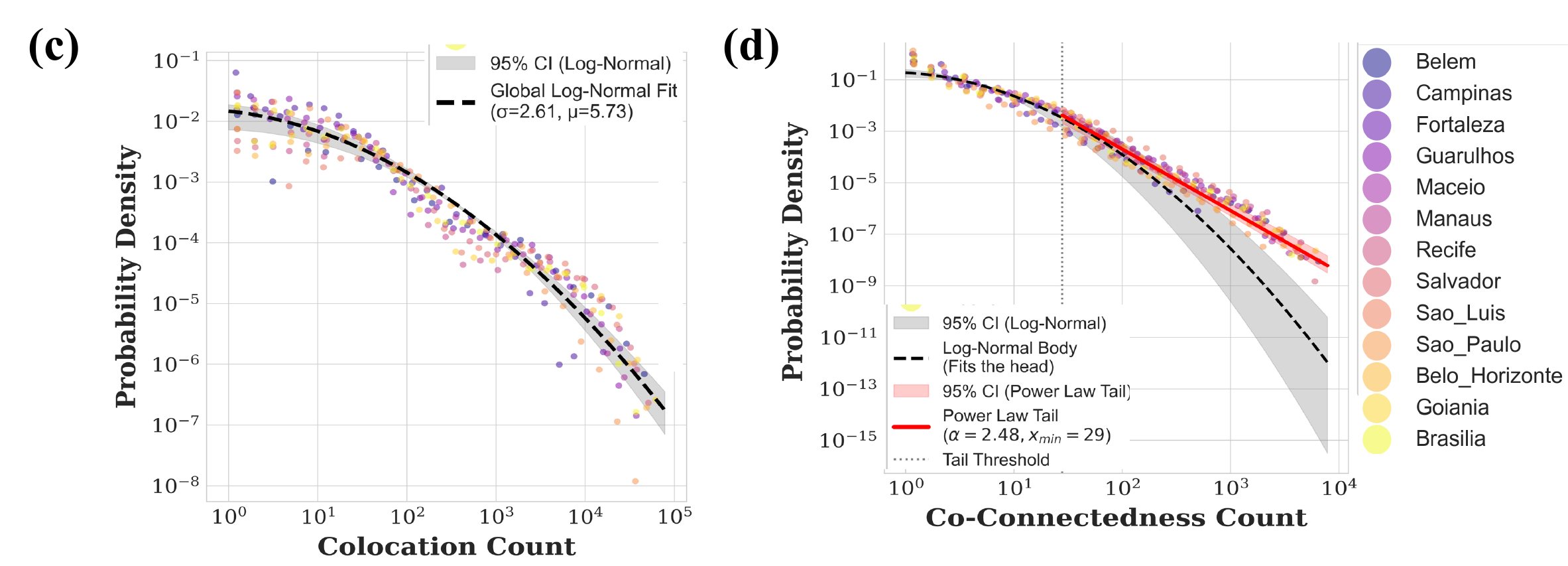}
    \caption{\textbf{Temporal sensitivity of colocation and co-connectedness distributions.} Probability density distributions of colocation ($C^A$) and co-connectedness ($O^{AB}$) obtained using co-presence thresholds of 15 minutes (top panels) and 45 minutes (bottom panels). The qualitative forms of both distributions are robust across temporal thresholds. Colocation remains broadly consistent with a log-normal form, while co-connectedness continues to exhibit a log-normal body with a heavy-tailed upper regime. These results indicate that the principal distributional properties of colocation and co-connectedness are robust to the choice of co-presence window.}
    \label{fig:figS3}
\end{figure}

\section{Iterated LouBar Procedure and Activity Levels ($L$)}\label{sec:supp_loubar}
\label{sec:loubar_levels}

To classify locations into the activity levels used to construct the interaction matrices and compute the hierarchy metric ($\Phi$), we implement an iterated LouBar procedure based on the empirical distribution of colocation ($C^A$). Rather than applying a single threshold, we recursively apply the LouBar geometric projection to the sorted colocation values. In the first iteration, nodes comprising the upper tail (values above the LouBar tangent intercept) are assigned to Level 1 and removed from the dataset. The Lorenz curve is then recomputed for the remaining nodes, generating a new threshold for Level 2. This recursive separation continues until fewer than five nodes remain in the subset, or the mean of the remaining distribution reaches zero.

Consequently, the total number of activity levels, $L$, is not fixed globally but emerges from each city's empirical colocation distribution. Because the hierarchy metric ($\Phi$) can depend on matrix dimensionality, each empirical network is evaluated against Node Shuffling and Edge Shuffling null models constructed using the same activity-level partition and therefore the same value of $L$. These matched null models provide city-specific baselines for determining whether the observed concentration of co-connectedness within the same or neighboring activity levels exceeds that expected from the corresponding randomized organization.

\section{Null Models}\label{section:null_model}

\begin{figure}[htpb]
    \centering
    \includegraphics[width=\textwidth]{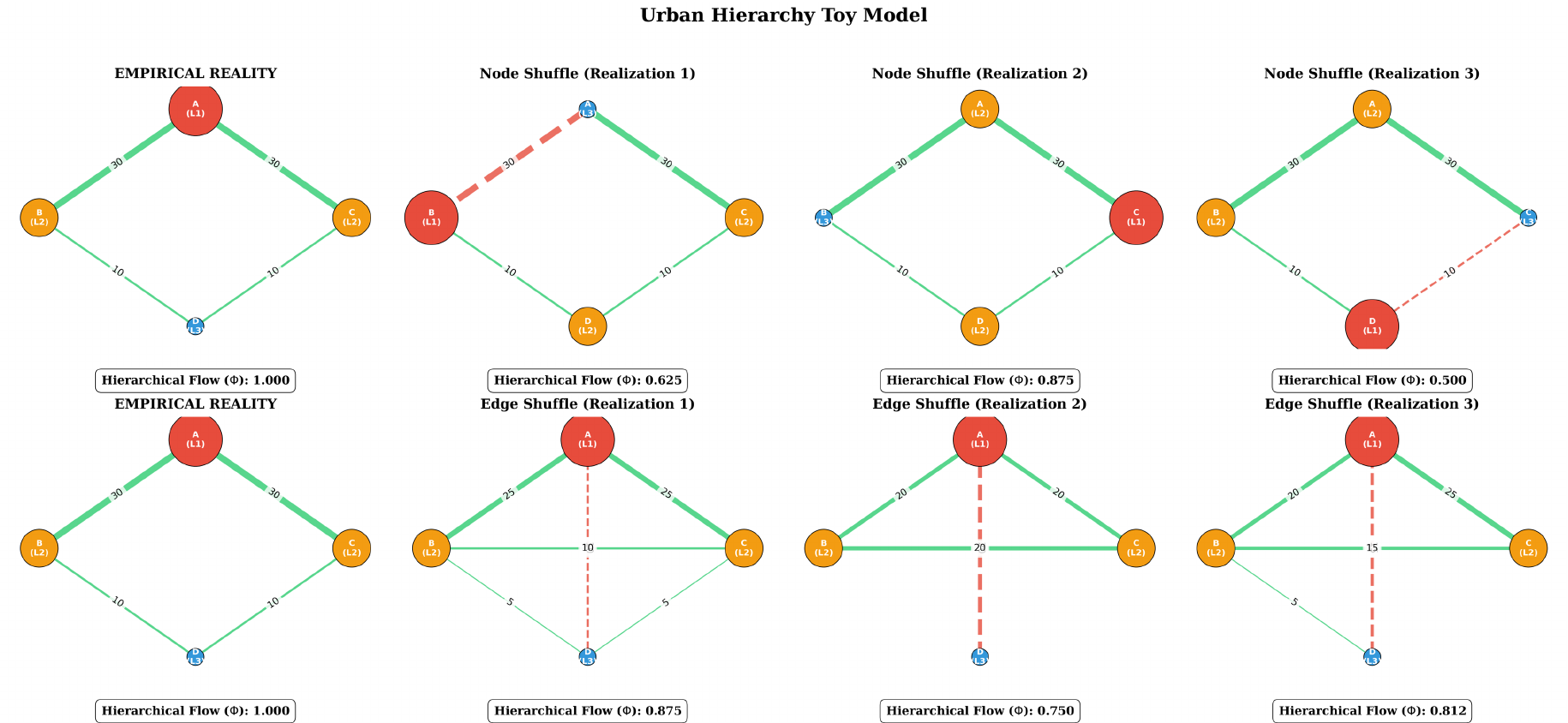}
    \caption{\textbf{Conceptual framework of the null models.} Illustration of the two randomization strategies used to evaluate hierarchical organization. Node Shuffling preserves the empirical co-connectedness network while randomizing the assignment of activity levels to locations. Edge Shuffling preserves the heterogeneous distribution of overlap across activity levels while randomizing the specific organization of co-connectedness between levels according to the heterogeneous-mixing null model. Together, the two models test whether the observed hierarchy depends on the correspondence between location activity and network structure and whether it exceeds that expected from activity heterogeneity alone.}
    \label{fig:FigS4}
\end{figure}

A high value of the hierarchy metric ($\Phi$) does not by itself imply non-random hierarchical organization. Some degree of concentration between similar activity levels may arise from heterogeneity in colocation activity and overlap volumes rather than from the specific organization of co-connectedness between levels. To evaluate whether the observed hierarchy exceeds these baseline effects, we compare the empirical networks against two complementary null models: Node Shuffling and Edge Shuffling (~\ref{fig:FigS4}).

\subsection{Node Shuffling}\label{sec:supp_node_shuffle}

The Node Shuffling model preserves the empirical co-connectedness network while randomizing the assignment of activity levels to locations. As a result, the network topology and edge weights remain unchanged, but the correspondence between location activity levels and the structure of the co-connectedness network is removed.

This model tests whether the observed hierarchy depends on this correspondence rather than arising solely from the underlying co-connectedness network. As shown in the main text, randomizing activity-level assignments substantially reduces $\Phi$, indicating that the observed diagonal concentration depends on the association between location activity levels and the pattern of co-connectedness between locations.

\subsection{Edge Shuffling}\label{sec:supp_edge_shuffle}

The Edge Shuffling model provides a more stringent baseline by preserving the heterogeneous distribution of overlap across activity levels while removing the empirical organization of co-connectedness between levels. Following the heterogeneous-flow null model introduced for mobility networks in Ref.~\hyperlink{SIref2}{[2]}, the expected interaction between levels $i$ and $j$ is given by

\begin{equation}
T_{ij}^{h} =
\sum_{k=1}^{L} T_{ik}
\frac{\sum_{m=1}^{L} T_{mj}}
{\sum_{m,k=1}^{L} T_{mk}}.
\end{equation}

Here, $\sum_k T_{ik}$ is the total co-connectedness associated with level $i$, while $\sum_m T_{mj}/\sum_{m,k}T_{mk}$ is the fraction of total co-connectedness associated with level $j$. The resulting matrix $T^h$ therefore represents the expected interaction between activity levels under random mixing conditioned on the empirical marginal overlap volumes.

To construct the null ensemble, we generate randomized realizations by rewiring connections according to the mixing probabilities implied by $T^h$, while preserving the corresponding empirical marginal constraints. This procedure retains the heterogeneous distribution of overlap across activity levels but removes the specific level-to-level organization observed in the empirical network. The Edge Shuffling model therefore tests whether the observed hierarchy exceeds that expected from activity heterogeneity alone.

\subsection{Statistical Evaluation}\label{sec:supp_stats}

To quantify the deviation of the empirical hierarchy from the corresponding null expectations, we generate an ensemble of 100 independent randomized network realizations for each null model and each city. We calculate the mean ($\langle \Phi_{\mathrm{null}} \rangle$) and standard deviation ($\sigma_{\mathrm{null}}$) of the hierarchy metric across the resulting ensemble and compute the Z-score,

\[
Z=\frac{\Phi_{\mathrm{emp}}-\langle\Phi_{\mathrm{null}}\rangle}{\sigma_{\mathrm{null}}},
\]

where positive values indicate that the empirical hierarchy exceeds the mean hierarchy of the corresponding null ensemble. The magnitude of $Z$ measures this deviation in units of the standard deviation of the null distribution.

\subsection{Evaluating Structural Robustness: Node and Edge Shuffling}\label{sec:supp_robust_null}

\begin{figure}[htpb]
    \centering
    \includegraphics[width=\textwidth]{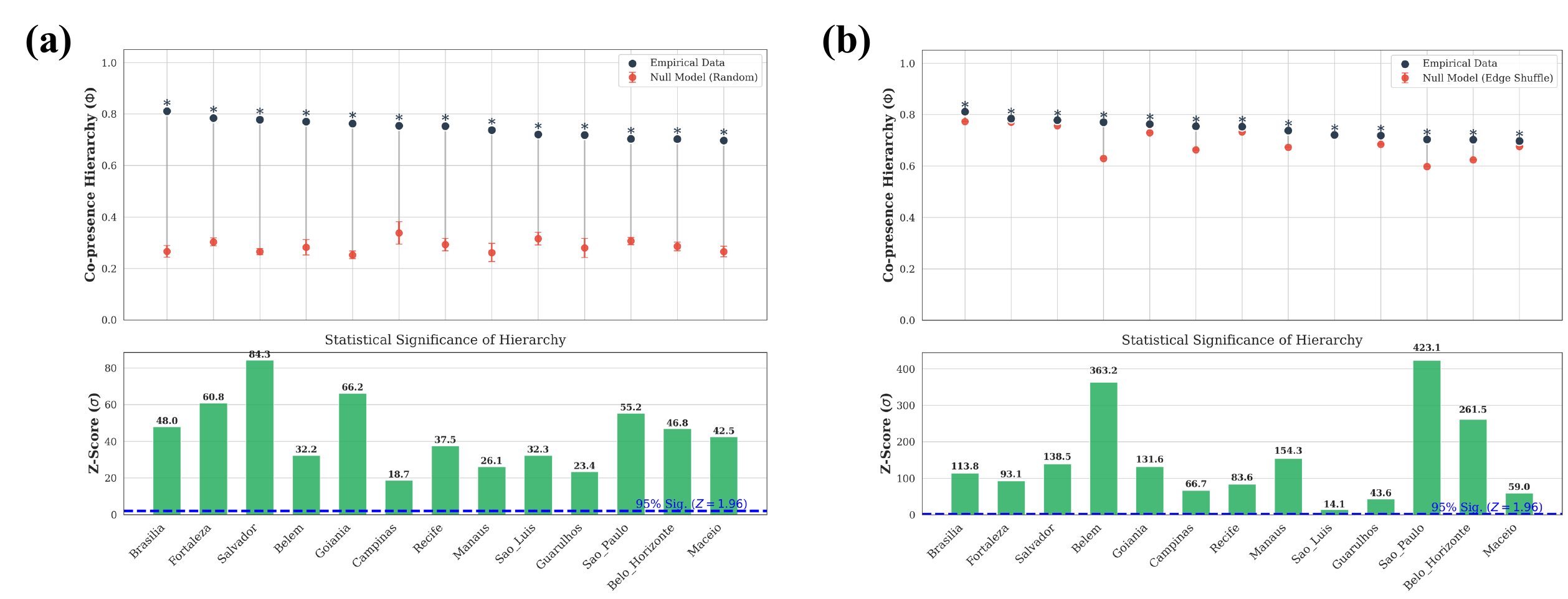}
    \caption{\textbf{Hierarchy of social overlap under complementary null models.}
    \textbf{(a) Node Shuffling.} The upper panel compares the empirical hierarchy ($\Phi$) with the distribution obtained by preserving the empirical co-connectedness network while randomizing the assignment of activity levels to locations. The lower panel shows the corresponding Z-scores calculated from 100 independent randomized realizations per city.
    \textbf{(b) Edge Shuffling.} The null model retains the heterogeneous distribution of overlap across activity levels while randomizing the specific organization of co-connectedness between levels. The resulting null hierarchy is closer to the empirical value than under Node Shuffling, but the empirical hierarchy remains above the null expectation in every city. The lower panel shows the corresponding Z-scores calculated from 100 independent randomized realizations per city.}
    \label{fig:FigS5}
\end{figure}

\begin{figure}[htpb]
    \centering
    \includegraphics[width=\textwidth]{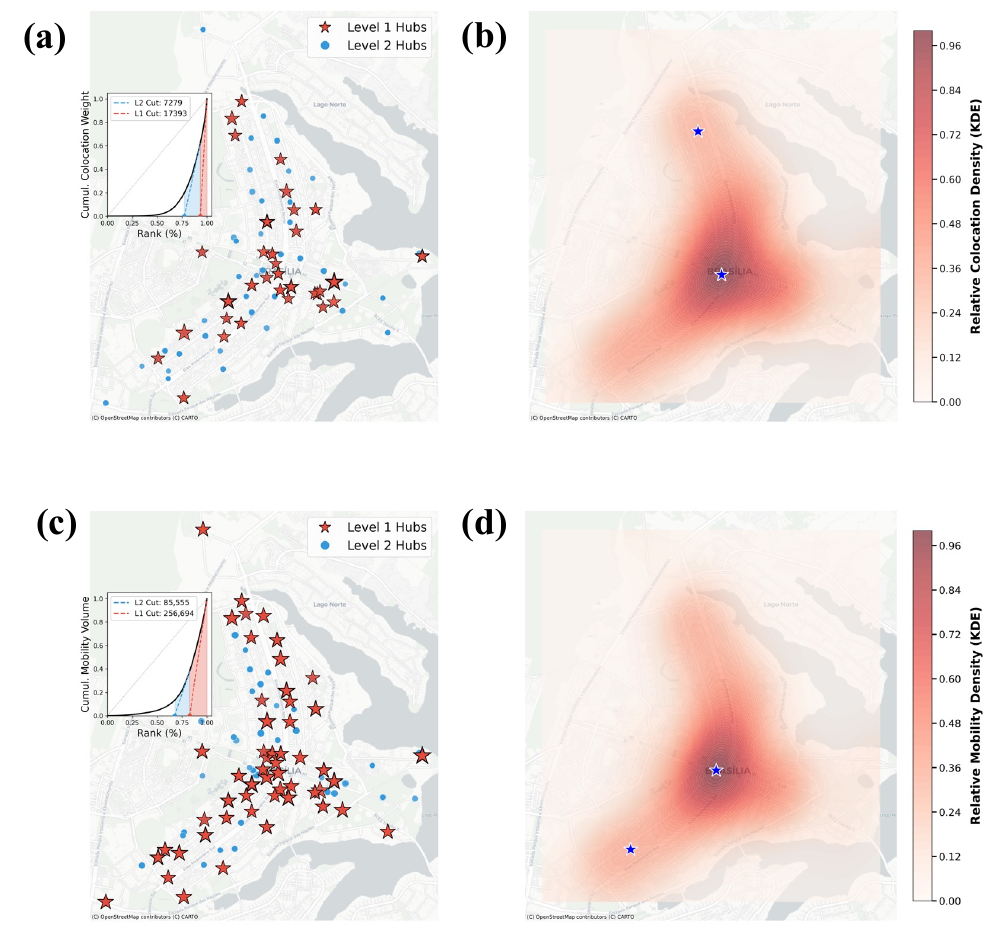}
    \caption{\textbf{Comparison of colocation and mobility centers in Bras\'ilia.}
    \textbf{(a)} Colocation hotspots identified from local colocation activity.
    \textbf{(b)} Social centers obtained after KDE smoothing and thresholding of colocation activity.
    \textbf{(c)} Mobility hotspots identified from movement flows.
    \textbf{(d)} Mobility centers obtained after KDE smoothing and thresholding of mobility activity.
    Although both representations identify two dominant centers, their spatial locations differ substantially.}
    \label{fig:FigS9}
\end{figure}

\begin{figure}[htpb]
    \centering
    \includegraphics[width=\textwidth]{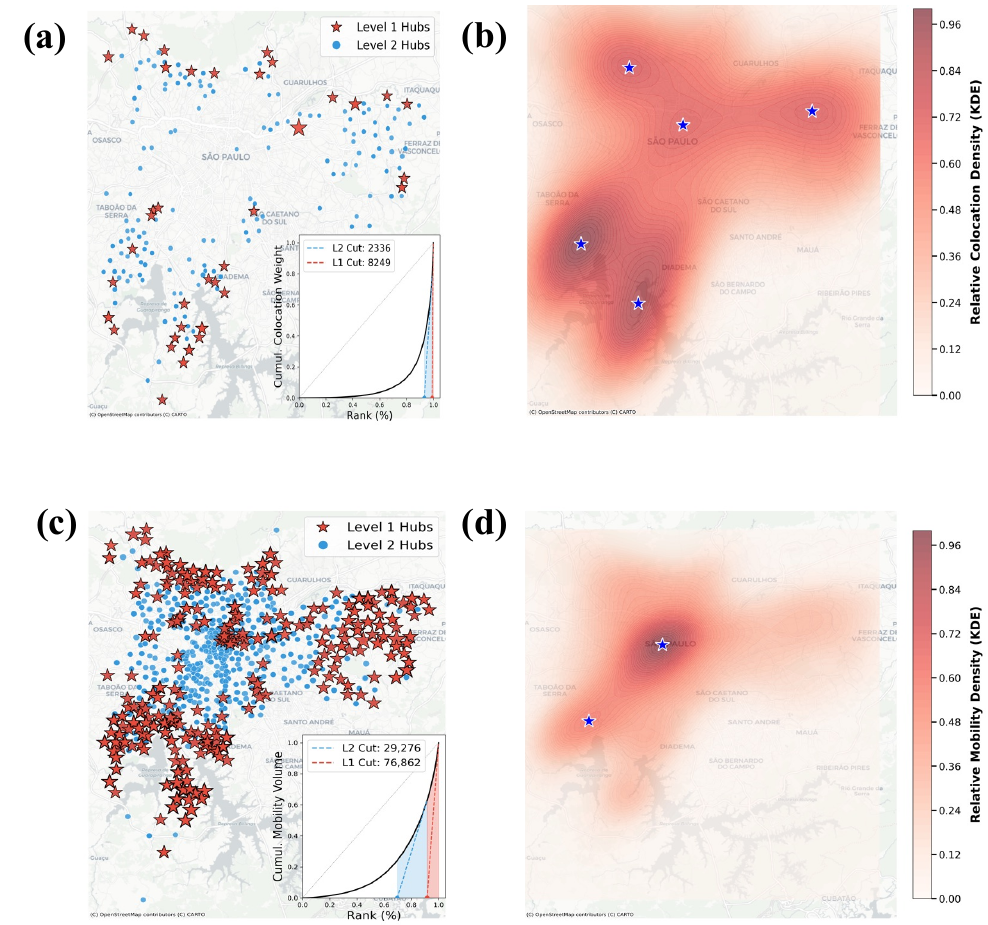}
    \caption{\textbf{Comparison of colocation and mobility centers in S\~ao Paulo.}
    \textbf{(a)} Colocation hotspots identified from local colocation activity.
    \textbf{(b)} Social centers obtained after KDE smoothing and thresholding of colocation activity.
    \textbf{(c)} Mobility hotspots identified from movement flows.
    \textbf{(d)} Mobility centers obtained after KDE smoothing and thresholding of mobility activity.
    The colocation representation identifies five dominant centers, whereas the mobility representation identifies two, illustrating differences in both centricity and spatial organization.}
    \label{fig:FigS10}
\end{figure}

Figure~\ref{fig:FigS5} compares the empirical hierarchy with the distributions obtained from the Node Shuffling and Edge Shuffling null models. For each city and each null model, the deviation of the empirical hierarchy ($\Phi$) from the null expectation is quantified using the Z-score calculated from 100 independent randomized realizations.

Under Node Shuffling, randomizing the assignment of activity levels while preserving the empirical co-connectedness network substantially reduces the expected hierarchy. Across all thirteen cities, the empirical hierarchy lies well above the corresponding null expectation, indicating that the observed diagonal concentration depends on the correspondence between location activity levels and the structure of the co-connectedness network.

The Edge Shuffling model provides a stronger control for activity heterogeneity by retaining the heterogeneous distribution of overlap across activity levels while removing the specific organization of co-connectedness between levels. As expected, the resulting null hierarchy is closer to the empirical value than under Node Shuffling. Nevertheless, the empirical hierarchy remains above the Edge Shuffling expectation in every city. The variance across Edge Shuffling realizations is also small, resulting in large positive Z-scores despite the smaller absolute difference between the empirical and null values.

Together, the two null models show that the observed hierarchical organization is not reproduced either by random assignment of activity levels to the empirical co-connectedness network or by random mixing conditioned on the heterogeneous distribution of overlap across activity levels. The observed concentration of co-connectedness between the same and neighboring activity levels therefore reflects structure beyond these two baseline effects.

\section{Difference between Colocation and Mobility Centers}\label{sec:supp_centers}

To illustrate the differences between mobility and colocation centricity, we compare the spatial distribution of the corresponding centers in Bras\'ilia and S\~ao Paulo. Bras\'ilia provides an example in which both representations identify two dominant centers but place them in different locations, whereas S\~ao Paulo exhibits differences in both the number and spatial arrangement of centers.

\section{Geographic Embedding and Urban Form}\label{section:geographic_embedding}

\begin{figure}[t!]
    \centering
    \includegraphics[width=0.7\textwidth]{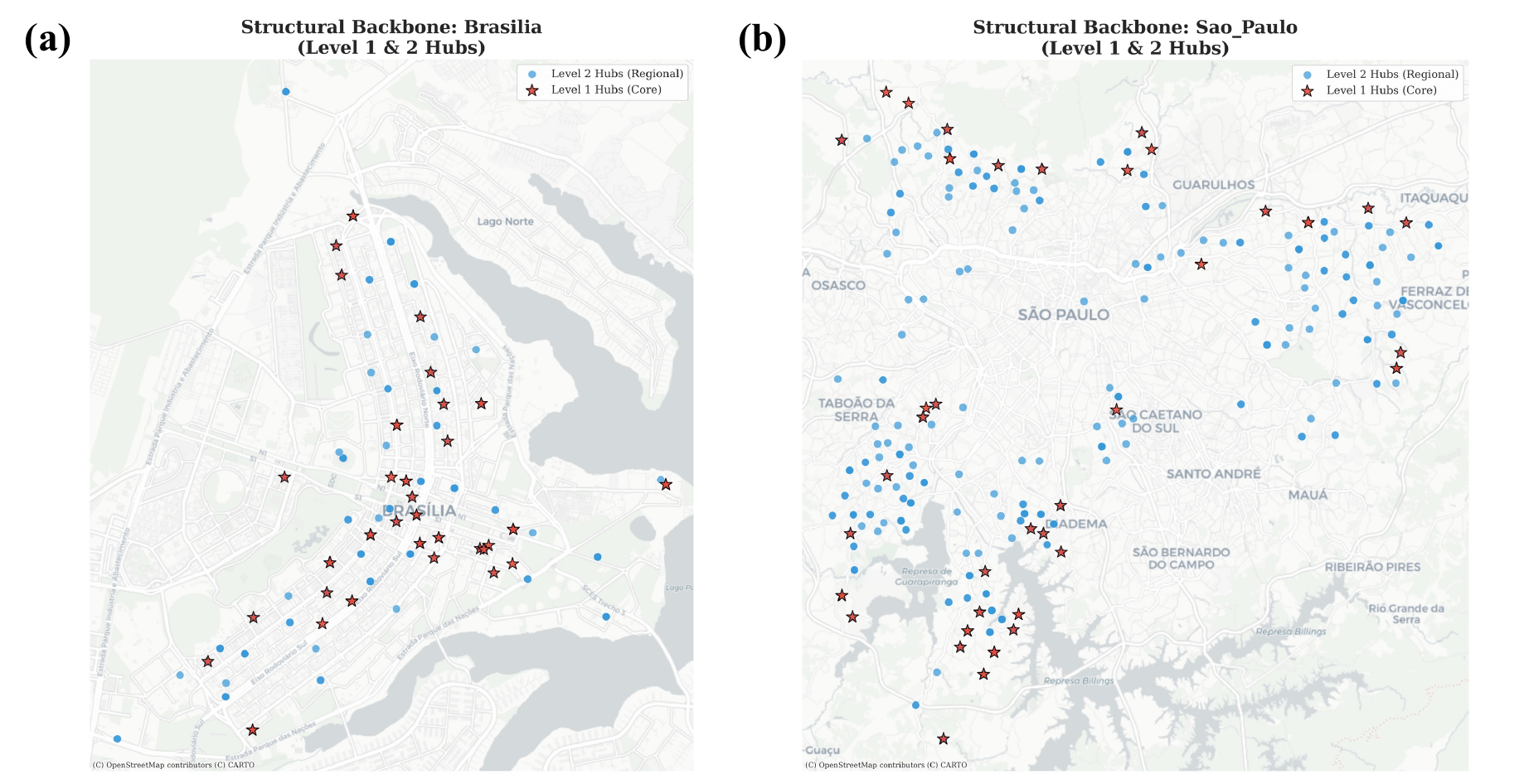}
    \caption{\textbf{Spatial distribution of Level 1 and Level 2 colocation locations.} Representative examples from Bras\'ilia \textbf{(a)} and S\~ao Paulo \textbf{(b)}. Although both cities contain multiple high-activity locations, their geographic arrangement differs substantially. Bras\'ilia exhibits a relatively concentrated configuration along a common urban axis, whereas the corresponding locations in S\~ao Paulo are distributed across a larger metropolitan area.}
    \label{fig:FigS6}
\end{figure}

The hierarchy metric quantifies how co-connectedness is distributed across activity levels, but it does not encode the geographic arrangement of the locations belonging to those levels. Figure~\ref{fig:FigS6} illustrates this distinction by showing the spatial distribution of Level 1 and Level 2 colocation locations in Bras\'ilia and S\~ao Paulo.

Although both cities contain multiple high-activity locations, their geographic organization differs substantially. In Bras\'ilia, Level 1 and Level 2 locations are relatively concentrated along a common urban axis, whereas in S\~ao Paulo they are distributed across a much larger metropolitan area. The spatial arrangement of high-colocation locations therefore provides information about urban form that is distinct from the hierarchical organization of co-connectedness measured by $\Phi$.

Figure~\ref{fig:FigS6} complements the interaction matrices presented in the main text by making this distinction explicit. The interaction matrices describe how recurrent social overlap is organized across activity levels, whereas the maps show where locations belonging to those activity levels are situated geographically.

\section{Spatial Consolidation and Center Detection}
\label{sec:center_detection}

To characterize the continuous spatial distribution of colocation activity from discrete CDR antenna locations, we apply a smoothing and peak-detection procedure. Because colocation volumes are associated with individual antenna locations, directly thresholding antenna-level activity can fragment spatially contiguous regions of high activity, particularly where antennas are densely distributed. We therefore construct a continuous spatial representation using Kernel Density Estimation (KDE), identify local maxima of the resulting density surface, and apply the LouBar thresholding procedure to distinguish dominant peaks. This procedure provides an operational definition of functional centers from the spatial distribution of colocation activity.

\subsection{Determining the Spatial Scale}\label{sec:supp_spatial_scale}

We first establish a characteristic spatial resolution for each city. Geographic coordinates of all active CDR antennas (WGS84) are transformed to the projected coordinate reference system EPSG:3857. We then calculate the distance from each antenna to its nearest neighboring antenna using a $k$-Nearest Neighbors (KNN) algorithm. Distances below 1 meter are excluded to avoid treating multiple antennas located at effectively the same coordinates as distinct spatial separations. The characteristic spatial resolution of the city, $d_{\mathrm{med}}$, is defined as the median of the remaining nearest-neighbor distances.

\subsection{Continuous Kernel Density Estimation (KDE)}\label{sec:supp_kde}

We estimate a continuous spatial field of colocation activity from the antenna-level observations using a two-dimensional Gaussian Kernel Density Estimation (KDE). Antenna locations define the spatial coordinates of the observations, and their corresponding colocation volumes provide the weights. The KDE is evaluated over a spatial grid spanning the geographic extent of each city, with grid resolution determined relative to the characteristic antenna spacing $d_{\mathrm{med}}$. The smoothing bandwidth, $h$, is determined using Scott's Rule, allowing the smoothing scale to adapt to the spatial distribution of antenna locations. The resulting density field, $Z(x,y)$, is normalized to the interval $[0,1]$. This procedure smooths antenna-level colocation activity into a continuous spatial representation from which spatially coherent regions of high activity can be identified.

\subsection{Local Maximum Detection and the LouBar Procedure}\label{sec:supp_peak_loubar}

To identify candidate centers in the continuous colocation field, we extract local maxima from the KDE surface. A local-maximum filter is applied to the KDE grid using a spatial window defined by $d_{\mathrm{med}}$, yielding a set of candidate locations and their corresponding density amplitudes.

We then apply the LouBar procedure \hyperlink{SIref1}{[1]} to the distribution of peak amplitudes. The amplitudes are ranked and used to construct a Lorenz curve representing their cumulative contribution to the total density. Following the LouBar construction, a tangent is drawn to the Lorenz curve at its upper endpoint $(1,1)$. The slope of this tangent is

\[
S=\frac{Z_{\mathrm{max}}}{\mu},
\]

where $Z_{\mathrm{max}}$ is the maximum peak amplitude and $\mu$ is the mean amplitude across candidate peaks. The intersection of the tangent with the horizontal axis defines the cutoff

\begin{equation}
x_{\mathrm{cut}} = 1-\frac{1}{S}.
\end{equation}

This cutoff identifies the upper portion of the ranked peak distribution selected by the LouBar criterion. The corresponding peak amplitude defines the density threshold $Z_{\mathrm{cut}}$. Candidate local maxima with amplitudes exceeding this threshold, $Z(x,y)>Z_{\mathrm{cut}}$, are classified as functional social centers.

\subsection{Bandwidth Sensitivity Analysis}\label{sec:supp_bandwidth}

Because the numbers of colocation and residential centers ($N_{\mathrm{col}}$ and $N_{\mathrm{res}}$) depend on the Kernel Density Estimation (KDE) smoothing step, we evaluated their sensitivity to the bandwidth parameter. The primary analysis uses the bandwidth determined by Scott's Rule. As a sensitivity analysis, we repeated the center-detection procedure using bandwidths corresponding to $\pm20\%$ and $\pm50\%$ of this baseline value.

The inferred center counts were generally stable under moderate changes in bandwidth ($\pm20\%$). Larger perturbations produced the expected changes in spatial resolution: decreasing the bandwidth by $50\%$ generated a larger number of localized peaks, whereas increasing it by $50\%$ merged nearby peaks and produced fewer, more spatially aggregated centers. Thus, the principal center structure is robust to moderate variation around the bandwidth used in the primary analysis, while larger changes alter the spatial scale at which centers are resolved.

\section{Residential Population Assignment}
\label{sec:residential_centers}

To identify residential centers for comparison with the spatial distribution of colocation activity, we assign gridded population estimates to individual CDR antenna locations using Voronoi catchment areas.

\textbf{Defining Catchment Areas with Voronoi Polygons.} Using the coordinates of all active antennas, we construct a spatial boundary around each city from the convex hull of the antenna locations, extended by a $0.05^\circ$ buffer. The resulting study area is partitioned using a Voronoi tessellation, assigning each location within the boundary to its nearest antenna and thereby defining a geographic catchment area for each antenna.

\textbf{Integrating Grid-Based Population Data.} We overlay the resulting Voronoi polygons on high-resolution gridded population estimates from WorldPop (UN-adjusted). For each polygon, the population raster is cropped to the polygon boundary and missing or zero-valued pixels are excluded. The remaining pixel values are summed to estimate the residential population associated with the corresponding antenna catchment area.

\textbf{Finding Residential Centers.} Using the resulting residential population weights at antenna locations, we apply the same spatial procedure described in Section~\ref{sec:center_detection}: continuous KDE, local-maximum extraction, and LouBar thresholding. Applying the same center-detection procedure to colocation and residential population weights yields two sets of spatial centers: functional social centers derived from colocation activity and residential centers derived from population density. These centers are subsequently used to quantify the spatial proximity of colocation activity to social centers relative to residential centers.

\clearpage
\section*{Supplementary References}\label{sec:supp_refs}

\noindent\hypertarget{SIref1}{[1]} Louail, T., Lenormand, M., Cantu Ros, O. G., Picornell, M., Herranz, R., Frias-Martinez, E., Ramasco, J. J., \& Barthelemy, M. (2014). From mobile phone data to the spatial structure of cities. \textit{Scientific reports}, \textbf{4}(1), 5276.

\medskip
\noindent\hypertarget{SIref2}{[2]} Bassolas, A., Barbosa-Filho, H., Dickinson, B., Dotiwalla, X., Eastham, P., Gallotti, R., Ghoshal, G., Gipson, B., Hazarie, S. A., Kautz, H., et al. (2019). Hierarchical organization of urban mobility and its connection with city livability. \textit{Nature Communications}, \textbf{10}, 4817.

\end{document}